\documentclass[twocolumn, prd, nofootinbib]{revtex4}
\usepackage{amsmath,hyperref,graphicx,slashed,mathtools,color}

\usepackage{comment}

\newcommand{\K}{\textbf{K}}
\newcommand{\E}{\textbf{E}}

\newcommand{\inv}[1]{\frac{1}{#1}}

\begin{document}
\title{Ginzburg-Landau Type Approach to the 1+1 Gross Neveu Model - Beyond Lowest Non-Trivial Order}
\author{Anees Ahmed}
\email{anees.ahmed@uconn.edu}
\affiliation{Department of Physics, University of Connecticut, Storrs CT 06269, USA}
\begin{abstract}
This paper presents a case study of the effects of increasing the order of a Ginzburg-Landau type expansion, by using the well known Gross-Neveu model in 1+1 dimensions as a test case. It is found that as the order of expansion increases, the predicted phase diagram increasingly resembles the known exact phase diagram. Finally, some properties of arbitrary large order phase diagrams are examined.
\end{abstract}

\maketitle

\section{Introduction}
The Gross-Neveu model is a QFT with $N_f$ flavors of massless fermions with a four-fermion interaction term \cite{grossNeveu}:
\begin{equation} \label{eq:GNLag}
	\mathcal{L} = \bar{\psi_a} i  \slashed{\partial} \psi_a + \frac{g^2 }{2}  (\bar{\psi_a}\psi_a)^2, \text{\quad } a = 1,2,.., N_f
\end{equation}
Despite its simplicity, the 1+1 dimension version of the theory (referred to as GN$_2$ from here on) possesses several interesting properties in the limit $N_f \rightarrow \infty$. The theory shares several important features with Quantum Chromodynamics (QCD), such as asymptotic freedom, renormalizability and dynamical breakdown of the discrete chiral symmetry ($\psi \rightarrow \gamma_5 \psi$).  The latter is responsible for giving a dynamical mass to the otherwise massless fermions. 

The Lagrangian of the theory can be rewritten in terms of an auxiliary field $\phi$ by means of a Hubbard-Stratanovich transformation
\begin{equation} \label{eq:GNEffLag}
	\mathcal{L} = \bar{\psi_a} \left( i  \slashed{\partial} - \phi \right) \psi_a -\frac{1}{2g^2} \phi^2.
\end{equation}
The auxiliary field $\phi$ satisfies a consistency condition in the limit $N_f \rightarrow \infty$: 
\begin{equation}
	\phi = -g^2 \langle \bar{\psi}_a \psi_a \rangle.
\end{equation}
This suggests a physical interpretation of $\phi$ as a bosonic condensate. As evident from the kinetic part of the Lagrangian, this condensate acts as a mass term and breaks the discrete chiral symmetry (under which $\bar{\psi} \psi \rightarrow -\bar{\psi} \psi$). The fermionic fields can be integrated out, since the Lagrangian (\ref{eq:GNEffLag}) is quadratic in $\psi$, yielding an effective action
\begin{equation}
	S_\text{eff}[\phi] = -\frac{1}{2g^2 N_f} \int d^2x \text{ }\phi^2 - i \ln \det \left(  i  \slashed{\partial} - \phi\right)
\end{equation}
In the limit $N_f \rightarrow \infty$ the theory is explored by solving a \emph{gap equation}, which is just the extremum condition for the effective action with respect to $\phi$ :
\begin{equation} \label{eq:gap}
 	\frac{\delta S_\text{eff}}{\delta \phi} = 0 \implies \frac{1}{g^2 N_f} \phi + i \frac{\delta}{\delta \phi} \ln \det \left(  i  \slashed{\partial} - \phi\right) =0.
\end{equation}
For equilibrium thermodynamics, only static condenstates are considered: $\phi = \phi(x)$. Early work on the phase diagram of the model \cite{wolff} assumed a spatially homogeneous condensate, i.e. $\partial_x \phi = 0$. This assumption simplifies what is otherwise a much harder problem to solve, but misses several important features of the correct phase diagram. It was comparatively recently that the phase diagram was worked out without resorting to this assumption \cite{thiesOriginal,bdt}. In \cite{thiesOriginal}, the phase diagram is obtained via a relativistic Hartree-Fock approach.In \cite{bdt} the phase diagram is worked out by directly solving the gap equation \ref{eq:gap}. These two approaches are equivalent.  See \cite{thies-review} for a review.  This paper borrows results and notation from \cite{bdt} with minor alterations.

\begin{figure*} 
		\centering				
		\includegraphics[scale=1]{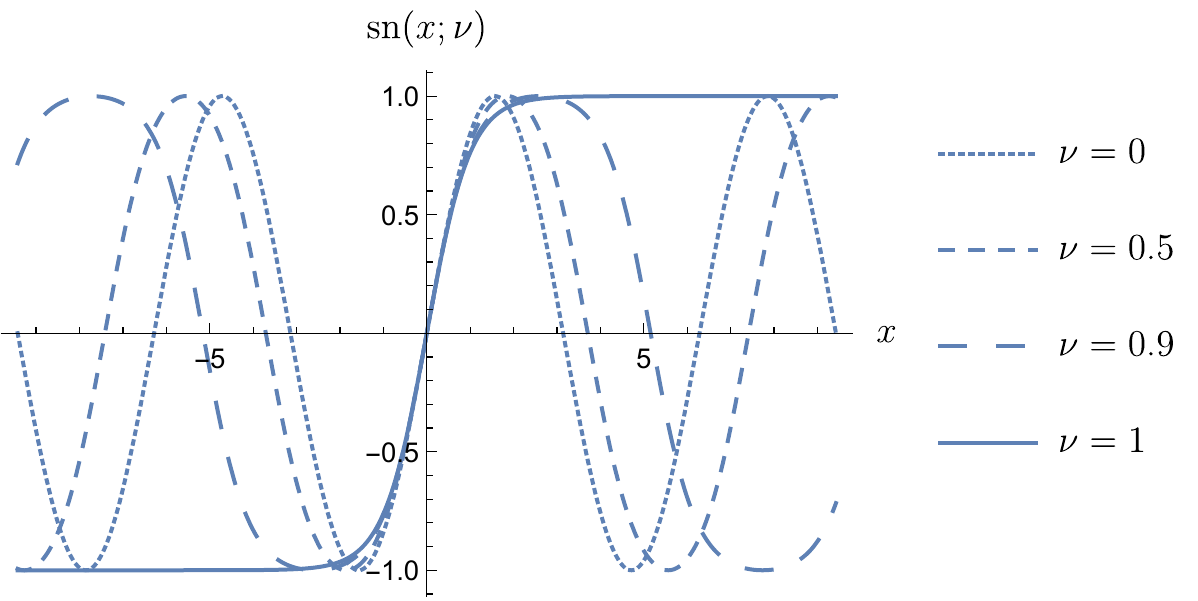}
		\caption{The Jacobi sine function sn($x;\nu$) for several values of the elliptic parameter $\nu$. The period increases with $\nu$ and at $\nu=1$ the function loses its periodicity and tends to a constant for large values of $|x|$.}
		\label{fig:jacobi}	
\end{figure*}

The most general solution to the gap equation \eqref{eq:gap} is 
\begin{equation} \label{condensateSolution}
		\phi(x; \lambda,\nu) = \lambda \sqrt{\nu} \text{ sn}(\lambda x;\nu).
\end{equation}
where sn is the Jacobi elliptic sine. The scale parameter $\lambda$ sets the size/amplitude of the condensate, while the elliptic parameter $\nu$ governs the shape of the condensate. See Fig. \ref{fig:jacobi}. The fact that the solution of the gap equation depends on only two parameters $\lambda$ and $\nu$ is remarkable, given the fact that the gap equation is a \emph{functional} equation. The existence of such a simple solution is a result of certain integrability properties of the GN$_2$ gap equation \cite{gokceOrig,belokolos}. 

The problem of solving the gap equation can be framed in an alternate way. In the semi-classical regime, the per-flavor grand canonical potential (GCP) $\Psi$ is just the effective action $S_\text{eff}[\phi(x)]$ times the temperature, with the condensate constrained to being static. Thus, solving the gap equation \eqref{eq:gap} is equivalent to minimizing the GCP with respect to $\phi(x)$. The GCP, in terms of the density of states $\rho(E)$, is given by
\begin{equation}
\begin{aligned}
 	\Psi[T,\mu;\phi(x)] = -T  \int_{-\infty}^\infty & dE \text{ } \rho(E) \ln\left( 1 + e^{-(E-\mu)/T} \right) \\ + & \frac{1}{2 N_f \, g^2 P} \int_0^P dx \, \left(\phi(x)\right)^2
\end{aligned}
\end{equation} 
where $P$ is the period of the condensate $\phi(x)$ \footnote{For $\phi$ given by \eqref{condensateSolution} the period is $4 \K(\nu)/\lambda$.}. This can be renormalized to \cite{grossNeveu,dashenGrossNeveu,thiesOriginal,bdt}
\begin{equation} \label{eq:GCP}
\begin{aligned}
	& \Psi[T,\mu;\phi(x)] = -T  \int_{-\infty}^\infty dE \text{ } \rho(E)  \ln\left( 1 + e^{-|E-\mu|/T} \right) \\
	 + &\int_{-\frac{\Lambda}{2} - \frac{Z(\nu)}{\Lambda}}^\mu dE \text{ } \rho(E) (E - \mu)  + \frac{\Lambda^2}{8\pi} + \frac{\Lambda\mu}{2\pi} +  \frac{Z(\nu)}{2\pi} \ln \Lambda.
\end{aligned}
\end{equation}
Here
\begin{equation}
		Z(\nu) = \frac{4}{\left( 1 + \sqrt{\nu} \right)^2}  \frac{1}{ P} \int_0^P dx \, \left(\phi(x)\right)^2
\end{equation}
and $\Lambda$ is the UV cut-off. The GCP depends on the condensate $\phi(x)$ through the density of states $\rho$, which for the solution \eqref{condensateSolution} takes the form
\begin{equation} \label{eq:DOS}
	\rho(E) = \frac{1}{ \pi} \frac{4E^2 + \lambda^2 \left(1 - \nu - 2 \frac{\E(\nu)}{\K(\nu)} \right)}{\sqrt{\left( 4 E^2 - \lambda^2 \left( 1+\sqrt{\nu}\right)^2 \right)\left( 4 E^2 - \lambda^2 \left( 1-\sqrt{\nu}\right)^2 \right)}}.
\end{equation}
The functions $\K$ and $\E$ are the complete elliptic integrals of the first and second kind, respectively.

Due to the simplicity of the solution \eqref{condensateSolution}, the problem of minimizing $\Psi$ with respect to $\phi(x)$ reduces to a much easier one of minimization with respect to two parameters, $\lambda$ and $\nu$.
\begin{equation}
	\frac{\delta}{\delta \phi} \Psi[T,\mu;\phi(x)] = 0 \xRightarrow{\scriptstyle{\phi = \lambda \sqrt{\nu} \text{sn}(\lambda x,\nu)}} \
	\begin{array}{l}
	 \partial_\lambda \Psi[T,\mu;\lambda,\nu] = 0 \\ 
	 \partial_\nu \Psi[T,\mu;\lambda,\nu] = 0
	\end{array} 
\end{equation}
The minimizing values of these parameters govern the phase of the GN$_2$ baryonic matter at any given point in the $\mu$-$T$ plane. This association of phase with the values of $\lambda$ and $\nu$ follows from the properties of the Jacobi elliptic sine sn: 

\begin{itemize}
	\item $\lambda = 0$, $\nu=0$ \quad - \quad massless homogeneous phase, translationally symmetric
	\item $\lambda \neq 0$, $\nu=1$ \quad - \quad massive homogeneous phase, translationally symmetric
	\item $\lambda \neq 0$, $0 < \nu < 1$ \quad - \quad massive crystalline phase, translationally asymmetric
\end{itemize}

Another way to explore the GN$_2$ model is through a Ginzburg-Landau type expansion of the GCP \eqref{eq:GCP}, with the condensate $\phi(x)$ as the inhomogeneous order parameter. This approach has several advantages. First, this approach is purely analytical for low orders. This makes it possible to extract expressions for certain features in the phase diagram. Second, Ginzburg-Landau expansions are typically phenomenological ansatzes based on symmetries of the system under study. See, for example, \cite{bogdanov,robler}, where the authors predict skyrmions in magnetic crystals from a Ginzburg-Landau expansion constructed from the symmetries of these crystals. Similarly one can probe to a certain degree Gross-Neveu models in any dimension via phenomenological expansions \cite{nickel}. These expansions will have the general form
\begin{equation}
 	\Psi_\text{GL} = \sum_n c_n(T,\mu) F_n \left[\phi, \vec{\nabla} \phi, \dots \right]
\end{equation}

GN$_2$ is a rare case where one can generate a Ginzburg-Landau expansion from the underlying microscopic theory, and so the form of the terms $F_n$  and the associated coefficients $c_n$ can be calculated exactly. Also, thanks to the integrability properties of the GN$_2$ model mentioned previously, the Ginzburg-Landau equation $\frac{\delta}{\delta \phi} \Psi_\text{GL} = 0$ is solved by \eqref{condensateSolution} at every order. Thus, GN$_2$ can be viewed as a test model where phase diagrams obtained from Ginzburg-Landau expansions can be studied at various orders and compared against the exact phase diagram. As will be seen later, it is found that the phase diagram generated from even a low order expansion exhibits several crucial features of the exact phase diagram such as existence of a crystalline phase and a tri-critical point, along with expressions for some of these features that match perfectly with the exact phase diagram. The following subsection discusses the GN$_2$ Ginzburg-Landau expansion in more detail.

\subsection{GN$_2$ Ginzburg-Landau expansion}
The Ginzburg-Landau expansion of the renormalized grand potential $\Psi$ (\ref{eq:GCP}) for any arbitrary condensate $\phi$ is obtained by expanding \eqref{eq:GCP} in powers of $\phi$ and its spatial derivatives (which reside in $\rho(E)$) \cite{thies,bdt}:
\begin{equation} \label{eq:OPExpGCP}
\begin{aligned}
	\Psi_\text{GL} = \alpha_0(T,\mu) + \alpha_2(T,\mu) \langle\phi^2\rangle + \alpha_4(T,\mu)  \langle\phi^4 + \phi'^2\rangle \\
	 +  \alpha_6(T,\mu)  \langle\phi^6  + 5 \phi^2 \phi'^2  + \frac{1}{2} \phi''^2 \rangle+ \dots
\end{aligned}
\end{equation}
where prime implies spatial derivative and $\langle \dots \rangle$ stands for spatial average over one period of $\phi$. The spatial derivatives  in the expansion are responsible for the spatially inhomogeneous phase that is known to exist in the exact GN$_2$ phase diagram.

The usual method to analyze a Ginzburg-Landau type GCP is to solve a Ginzburg-Landau equation, which is the extremum condition for the GCP with respect to the order parameter (which, in the semi-classical regime, is the same as solving the gap equation order by order). In the present case, this means solving the functional equation $\frac{\delta}{\delta \phi} \Psi_\text{GL}= 0$ with $\Psi_\text{GL}$ truncated at some desired order. This appears to be a formidable problem, especially at higher orders, but it so happens that the solution \eqref{condensateSolution} satisfies the Ginzburg-Landau equation to every order. This has to do with the fact that the Ginzburg-Landau equations form a heirarchy of differential equations known as the modified Korteweg-de Vries (KdV) hierarchy \cite{gokceOrig,bdt}. So instead of solving a different differential equation at every order, one can simply evaluate the Ginzburg-Landau GCP \eqref{eq:OPExpGCP} on the condensate solution \eqref{condensateSolution} to obtain an expansion that depends on only two variational parameters $\lambda$ and $\nu$
\begin{equation} \label{eq:GLExpAlln}
\Psi_\text{GL}(T,\mu;\lambda,\nu) = \alpha_0(T,\mu) +  \frac{1}{2\pi}\sum_{n=1}^\infty \lambda^{2n} \alpha_{2n}(T,\mu) f_{2n}(\nu),
\end{equation}
and then minimize with respect to $\lambda$ and $\nu$. Note that the expansion is now a small $\lambda$ expansion. Also, the dependence on $\lambda$ and $\nu$ has cleanly separated. This follows from the general structure of expansion, which can be understood by simple dimensional analysis. Take, for example, the $\alpha_4$ term in \eqref{eq:OPExpGCP} where the $\phi$ dependence is $\langle\phi^4 + \phi'^2\rangle $.  Each $\phi$ contributes one $\lambda$ to the spatial average, and every derivative contributes another $\lambda$ to the spatial average, so that the terms $\langle\phi^4\rangle$ and $\langle\phi'^2\rangle$ are both order $\lambda^4$ quantities. In general, the $\alpha_{2n}$ term will have the form $\langle \phi^{2n} + \dots \rangle$ where $\dots$ contains terms with such combinations of $\phi$ and its derivatives that the dependence on $\lambda$ is exactly $\lambda^{2n}$.

The $\nu$ dependence from the spatial averages is collected into the functions $f_{2n}(\nu)$ which are normalized such that they take values in the range [0,1]. These functions increase monotonically with $\nu$ and have the generic form
\begin{equation} \label{eq:fDef}
	f_{2n}(\nu) = p^{(2n-2)}(\nu) + q^{(2n-2)}(\nu) \frac{\E(\nu)}{\K(\nu)}
\end{equation}
where $p^{(n)}(\nu)$ and $q^{(n)}(\nu)$ are polynomials of order $n$. Explicit expressions for a few $f_{2n}(\nu)$ are listed below:
\begin{equation}
	\begin{aligned}
	&f_2(\nu) = 1 - \frac{\E}{\K} \\
	&f_4(\nu) = \inv{3} \Bigg(1 + 2\nu - (1 + \nu)\frac{\E}{\K}\Bigg)	\\
	&f_6(\nu) = \inv{10} \Bigg( (1 + 6\nu + 3\nu^2) - (1 + 4\nu + \nu^2) \frac{\E}{\K} \Bigg) \\
	&f_8(\nu) = \inv{35} \Bigg( (1 + 12\nu + 18 \nu^2 + 4\nu^3) \\ & \qquad \qquad- (1 + 9\nu + 9\nu^2 + \nu^3) \frac{\E}{\K} \Bigg) \\
	&f_{10}(\nu) = \inv{126} \Bigg( (1 + 20\nu + 60 \nu^2 + 40\nu^3 + 5 \nu^4) \\ &\qquad \qquad- (1 + 16\nu + 36\nu^2 + 16\nu^3 + \nu^4) \frac{\E}{\K} \Bigg)
	\end{aligned}
\end{equation}
The appearance of the functions $f_{2n}(\nu)$ in the expansion is a result of terms with derivative of $\phi$ in the expansion \eqref{eq:OPExpGCP}. Without the derivative terms, the expansion reduces to that of a homogeneous condensate, and the resulting phase diagram lacks any inhomogeneous phases. In terms of the functions $f_{2n}(\nu)$ this means that to recover the expansion for a homogeneous condensate, one has to take the limit $\nu \rightarrow 1$, which sets all $f_{2n}(\nu)$ to 1 for all $n$. In fact, setting $\nu$ to 1 (and consequently $f_{2n}(\nu)$ to 1) in any of the equations \eqref{eq:GCP}-\eqref{eq:GLExpAlln} produces the corresponding quantities for a spatially homogeneous condensate.

The functions $\alpha_{2n}(T,\mu)$ are given by
\begin{equation} \label{eq:alphaExp}
		\begin{aligned}
		& \alpha_0(T,\mu) = -\frac{\pi T^2}{6} - \frac{\mu^2}{2 \pi}\\
		& \alpha_{2}(T,\mu) =  \text{Re } \psi\left(\frac{1}{2} + i\frac{\mu}{2\pi T} \right) +  \ln 4\pi T \\
		& \alpha_{2n}(T,\mu) = \frac{(-1)^{n-1} }{n!(n-1)!} \frac{1}{(4 \pi T )^{2n-2}  } \\
		 & \quad \qquad \qquad \times \text{Re } \psi^{(2n-2)}\left(\frac{1}{2} + i\frac{\mu}{2\pi T} \right) , \quad n \geq 2
		\end{aligned}
	\end{equation}
where $\psi^{(n)}$ is the polygamma function of order $n$. Polygamma functions with identical $\mu$ and $T$ dependence appear naturally in expansions of thermodynamical quantities of the free Fermi gas \cite{blinnikov}. It is through the log and quadratic $T$-dependence in $\alpha_0(T,\mu)$ and $\alpha_2(T,\mu)$ that the four-fermion interaction of the GN$_2$ model manifests itself. Algebraically, the log and quadratic dependences are a result of renormalisation. Higher $\alpha_{2n}(T,\mu)$, beginning from $n=2$, are unaffected by renormalization. A simple way to see this is by expanding \eqref{eq:DOS} about $\lambda=0$ and plugging into \eqref{eq:GCP}\footnote{This is, in fact, the quickest way to derive the small $\lambda$ expansion \eqref{eq:GLExpAlln}}. The result is a series with integrals of the form 
\begin{equation}
 	\int_{-\infty + i \epsilon}^{\infty + i \epsilon}  \frac{dE}{E^{2n}} \ln\left( 1 + e^{-(E-\mu)/T} \right).
 \end{equation}
The contour must be shifted to avoid the pole at $E=0$. For $n=0$ and 1, this integral is UV divergent. Regularization by introducing a UV cutoff produces the quadratic $T$ dependence in $\alpha_0$ and the log dependence in $\alpha_2$. For $n \geq 2$, the integral is convergent and so there is no need for regularization.  

\section{Phase diagrams from the Ginzburg-Landau expansion}
\begin{figure} [!t]			
		\includegraphics[scale=0.33]{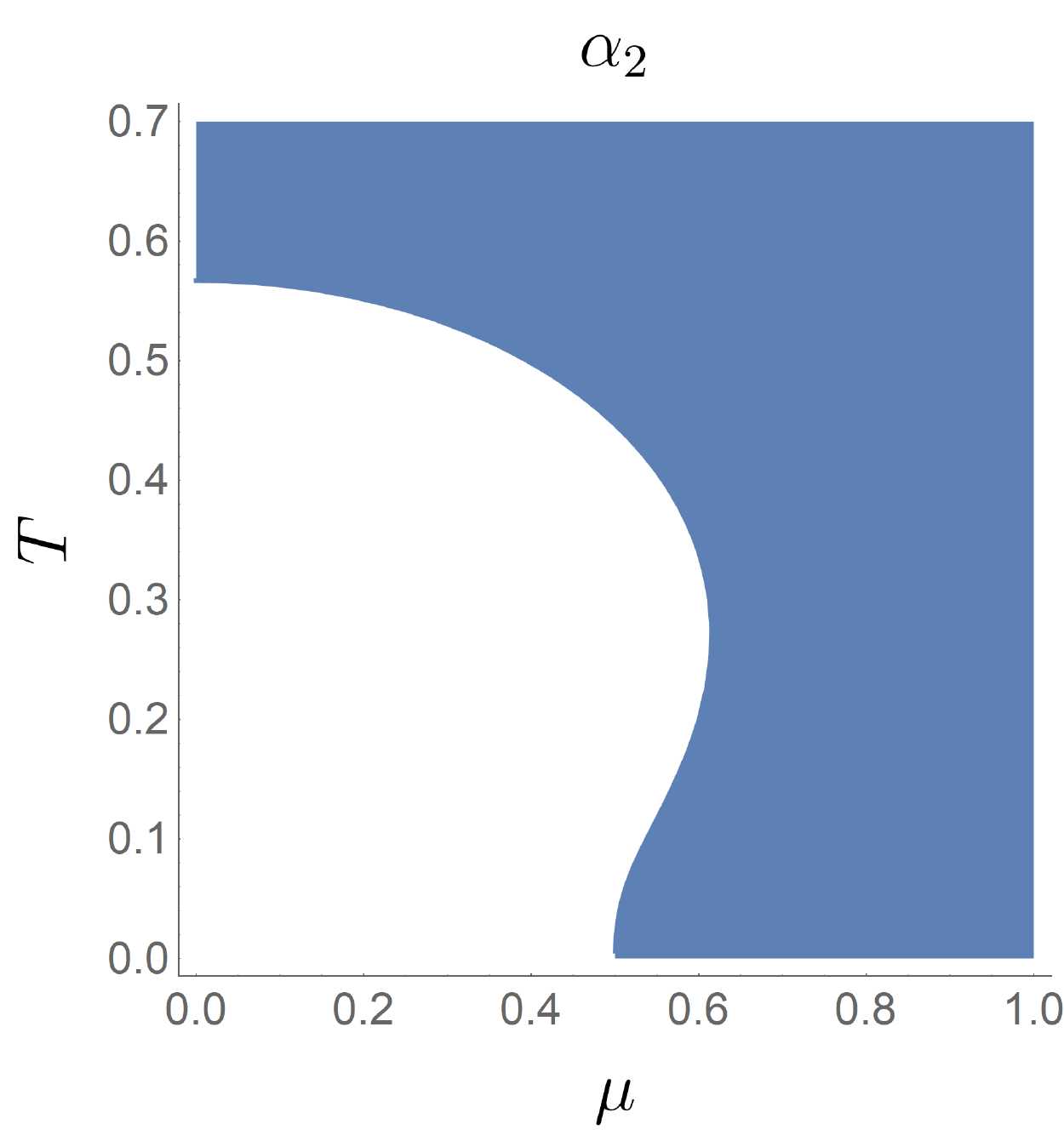}	\includegraphics[scale=0.33]{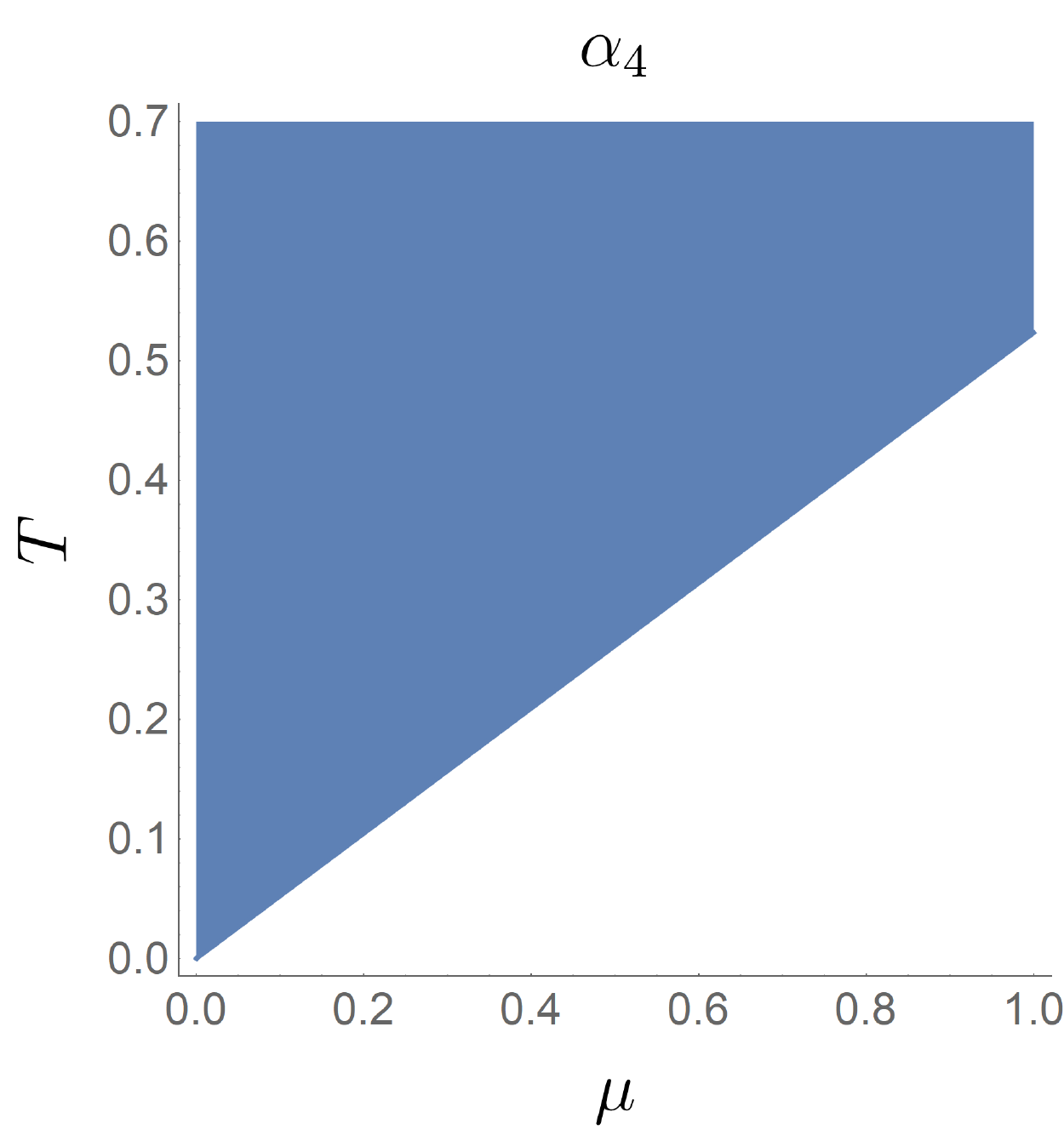}	\\
		\includegraphics[scale=0.33]{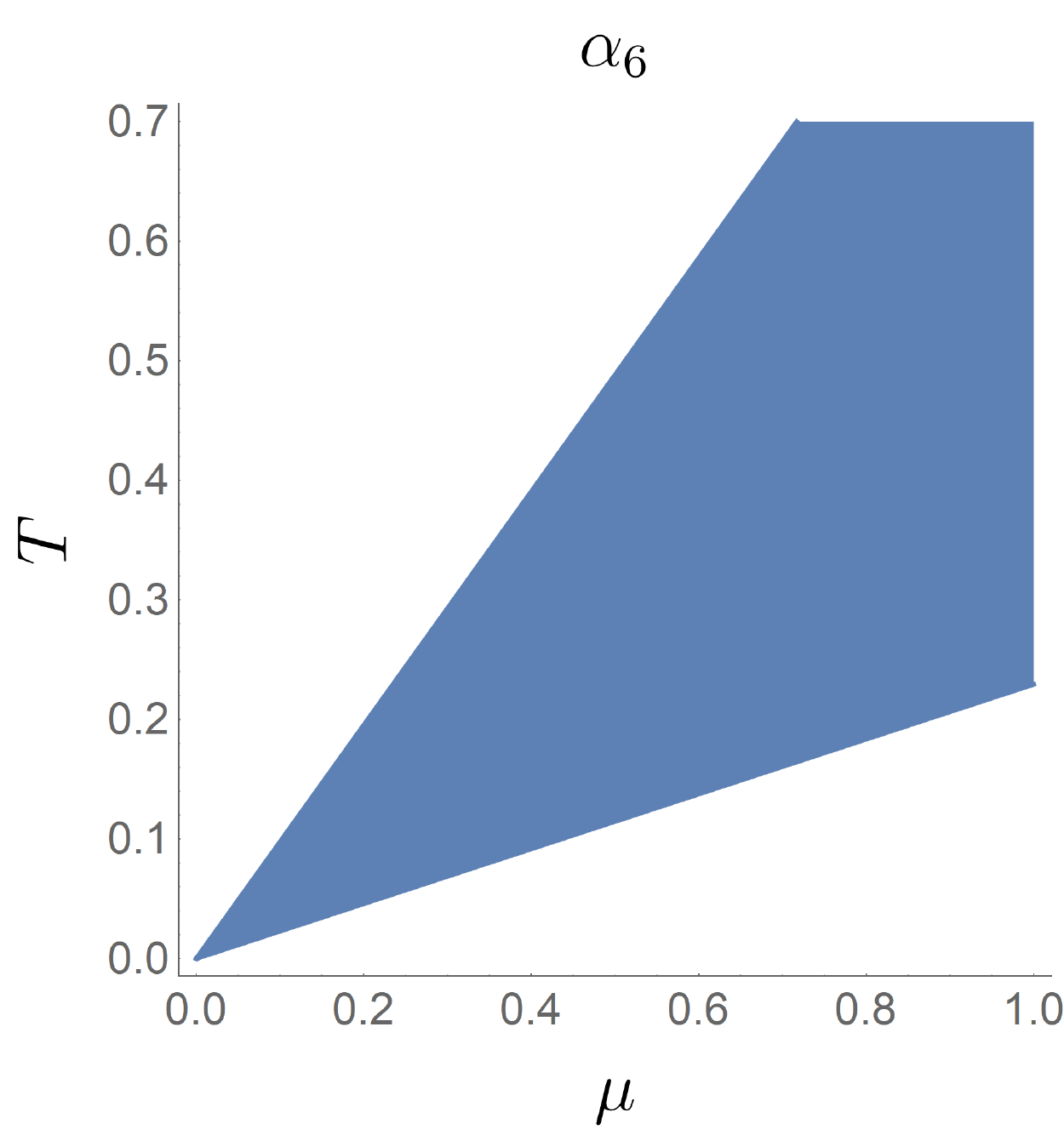}	\includegraphics[scale=0.33]{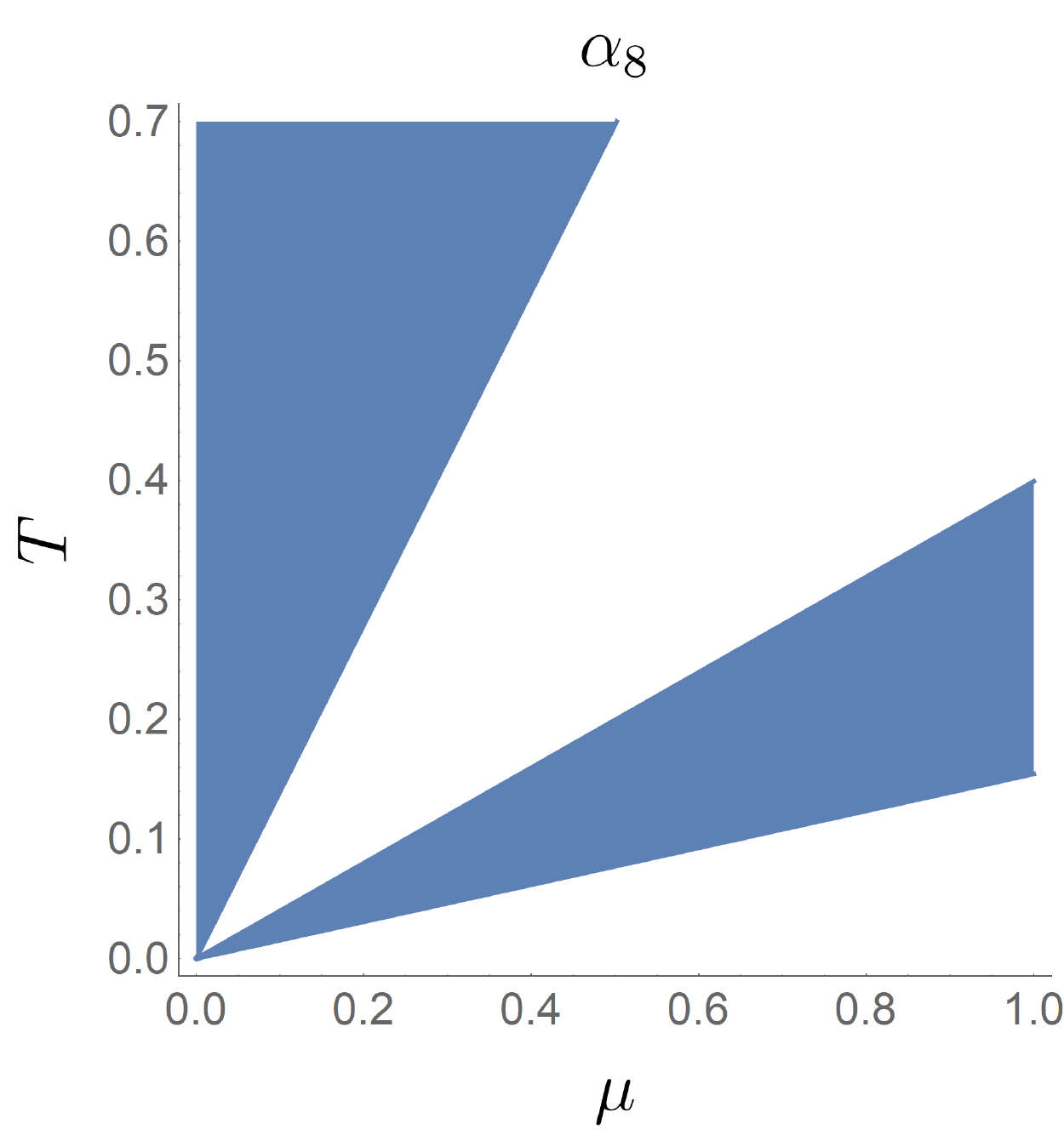} 	\\
		\includegraphics[scale=0.33]{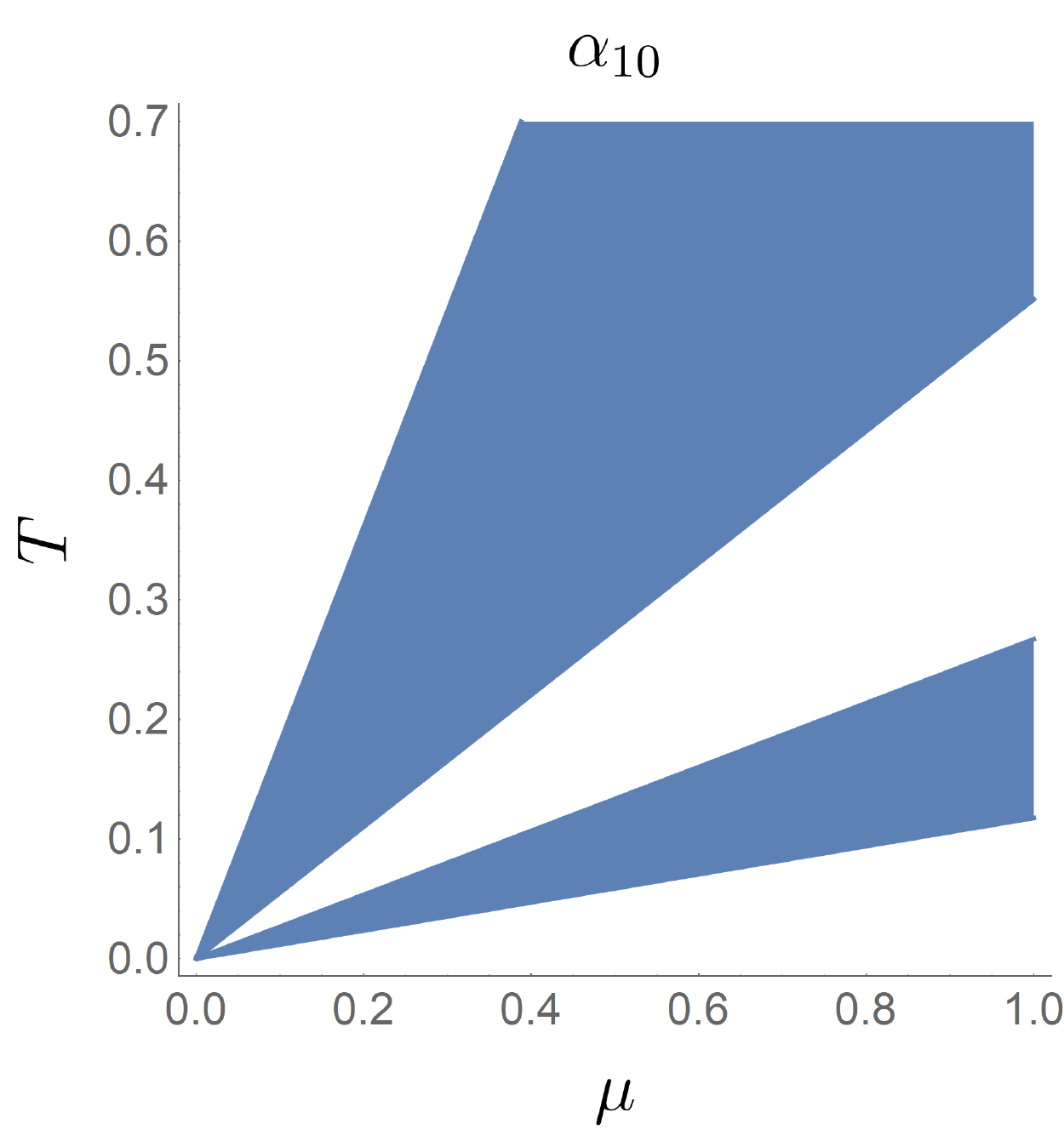} 	\includegraphics[scale=0.33]{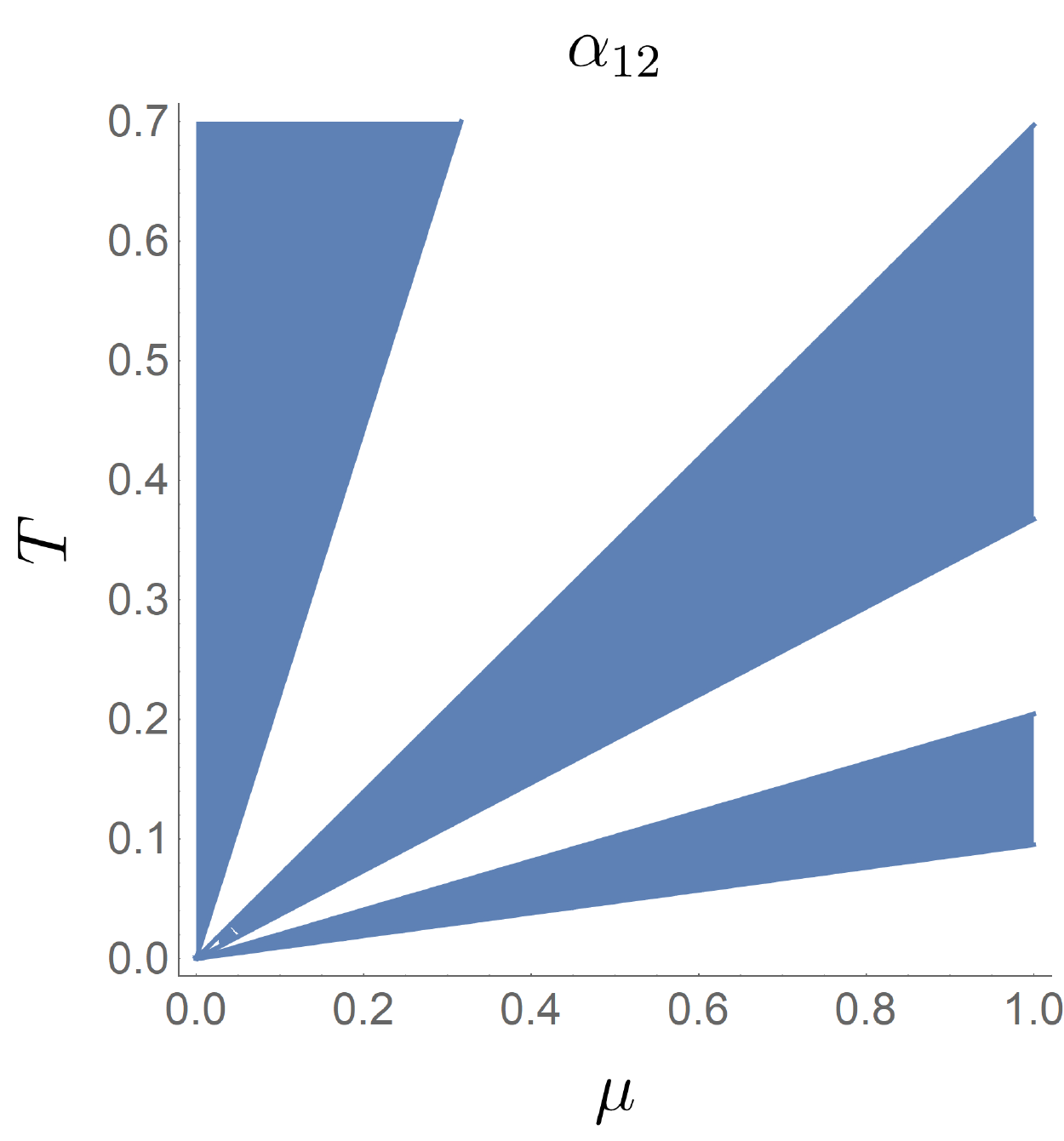}
		\caption{Regions in the $\mu-T$ plane where $\alpha_{2n}(T,\mu) > 0$, n = 1,2\dots 6. The number of wedges increases as $n$ increases. Also note that the slope of the lowest line on which $\alpha_{2n}(T,\mu)$ vanishes decreases with $n$.}
		\label{fig:SignsAlpha}	
	\end{figure}
This section explores phase diagrams obtained from Ginzburg-Landau expansion  \eqref{eq:GLExpAlln} at various orders. Let $\Psi^{(2N)}$ denote the expansion truncated (inclusive) at a finite order $\lambda^{2N}$, that is
\begin{equation} \label{eq:truncatedGCP}
	\Psi^{(2N)}(T,\mu;\lambda,\nu) = \alpha_0(T,\mu) +  \frac{1}{2\pi}\sum_{n=1}^N\lambda^{2n} \alpha_{2n}(T,\mu) f_{2n}(\nu)
\end{equation}
 Then $\Psi^{(2N)}$ is minimized with respect to $\lambda$ and $\nu$ at each point in the $\mu-T$ plane, that is  
\begin{equation} \label{eq:minCond}
		\partial_\nu \Psi^{(2N)} = 0 \qquad \text{and } \qquad \partial_\lambda \Psi^{(2N)}= 0
\end{equation}
are solved to obtain $\lambda(T,\mu)$ and $\nu(T,\mu)$, subject to the constraint that the Hessian matrix
\begin{equation} \label{eq:Hessian}
			\begin{pmatrix}
	\partial^2_\nu  \Psi^{(2N)} & \partial_\nu \partial_\lambda \Psi^{(2N)}\\
	\partial_\lambda \partial_\nu \Psi^{(2N)} & \partial^2_\lambda \Psi^{(2N)}
	\end{pmatrix}
\end{equation}
evaluated on these solutions is positive definite.
 Note that if all $\alpha_{2n}(T,\mu)$ functions are negative in a certain region of the $\mu-T$ plane, then there can be no minimum in that region. Such a region must be excluded from the analysis, as it is an artifact introduced by truncation. Further, because \eqref{eq:GLExpAlln} is a small $\lambda$ expansion, it is expected that the resulting phase diagram is accurate close to the massless condensate ($\lambda=0$) phase boundary, but the accuracy may fall off in regions distant from the massless phase boundary.

The grand potential can possibly have a global minimum whenever the highest order coefficient, $\alpha_{2N}(T,\mu)$, is positive; otherwise it is unbounded from below. Obviously, the grand potential can have local minima in either case. A phase diagram constructed based only on global minimum will be increasingly fragmented as the order of expansion $N$ increases, and will miss features that are exhibited by the exact phase diagram. To understand this, note that for $N>1$ the equation $\alpha_{2N}(T,\mu) = 0$ actually defines $N-1$ lines in the $\mu-T$ plane. Thus, as $N$ increases, the region where $\alpha_{2N}(T,\mu)$ is positive splits into multiple increasingly narrower wedges. See Fig. \ref{fig:SignsAlpha} for this behavior. Hence, in absence of a genuine global minimum, a local minimum must be chosen.

\subsection{Phase diagram to order $\lambda^4$}
\begin{figure} [!t]
		\centering				
		\raggedright \includegraphics[scale=0.7]{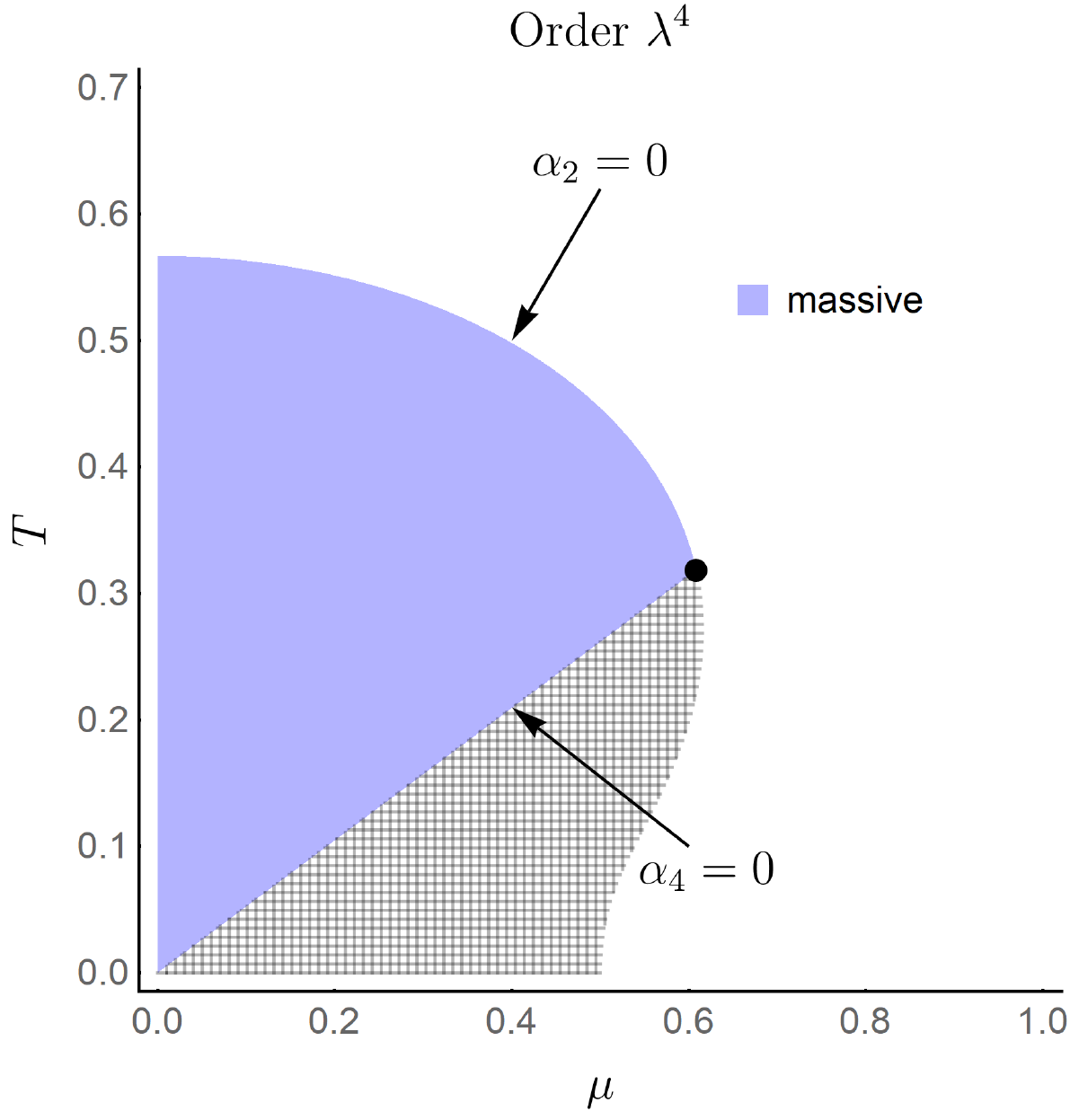}
		\caption{\textbf{Order $\lambda^4$ phase diagram}: Dot - tri-critical point, white - massless phase, mesh -  analysis is invalid because $\Psi^{(4)}$ is unbounded from below. The tri-critical point is actually not a feature of the phase diagram at order $\lambda^4$, but is included here for comparison and visualization purposes.}
		\label{fig:SecondOrderPhaseDiagram}	
\end{figure}
The GCP truncated at order $\lambda^2$ ($N=1$ in \eqref{eq:truncatedGCP}) has a minimum at $\lambda = 0$ in the region $\alpha_2(T,\mu) >0$, and none in $\alpha_2(T,\mu) < 0$. Thus at order $\lambda^2$, the phase diagram is trivial with just a single phase (massless homogeneous) and no phase transitions.

The first signs of any phase transitions are seen at order $\lambda^4$. At this order the minimization equations are
\begin{equation} 
	\begin{aligned} \label{eq:alpha4min}
		\partial_\lambda \Psi^{(4)}&= 0 \implies 2\lambda\left( \alpha_2 f_2 + 2 \lambda^2 \alpha_4 f_4 \right) = 0, \\
		\partial_\nu \Psi^{(4)} &= 0 \implies \lambda^2\left(\alpha_2 f_2' + \lambda^2 \alpha_4 f_4'\right) = 0.
	\end{aligned}
	\end{equation}
The only solution to these coupled equations which is actually a minimum is $\lambda = 0 $ in the region $\alpha_2(T,\mu) > 0$. This is the massless homogeneous phase. In addition, in the region $\alpha_2(T,\mu) <0$,  $\alpha_4(T,\mu) >0$ there is a minimum which is not local but global:
\begin{equation}
 	\lambda = \sqrt{- \frac{\alpha_2}{2 \alpha_4}}, \quad \nu = 1 \quad \quad   .
\end{equation}
This is the massive homogeneous phase. Thus, there are two phases at order $\lambda^4$ - massless homogeneous (in $\alpha_2(T,\mu) > 0$) and massive homogeneous (in $\alpha_2(T,\mu) <0, \alpha_4(T,\mu) >0$). The phase boundary between the two phases is given by the curve $\alpha_2(T,\mu) = 0$ constrained to the region $\alpha_4(T,\mu) > 0$. This phase boundary agrees perfectly with the exact phase diagram, and in fact does not change as the order of expansion \eqref{eq:truncatedGCP} is increased. See Fig. \ref{fig:SecondOrderPhaseDiagram} for the phase diagram.

\subsection{Phase diagram to order $\lambda^6$}
\begin{figure} [!t]
		\centering				
		\raggedright \includegraphics[scale=0.7]{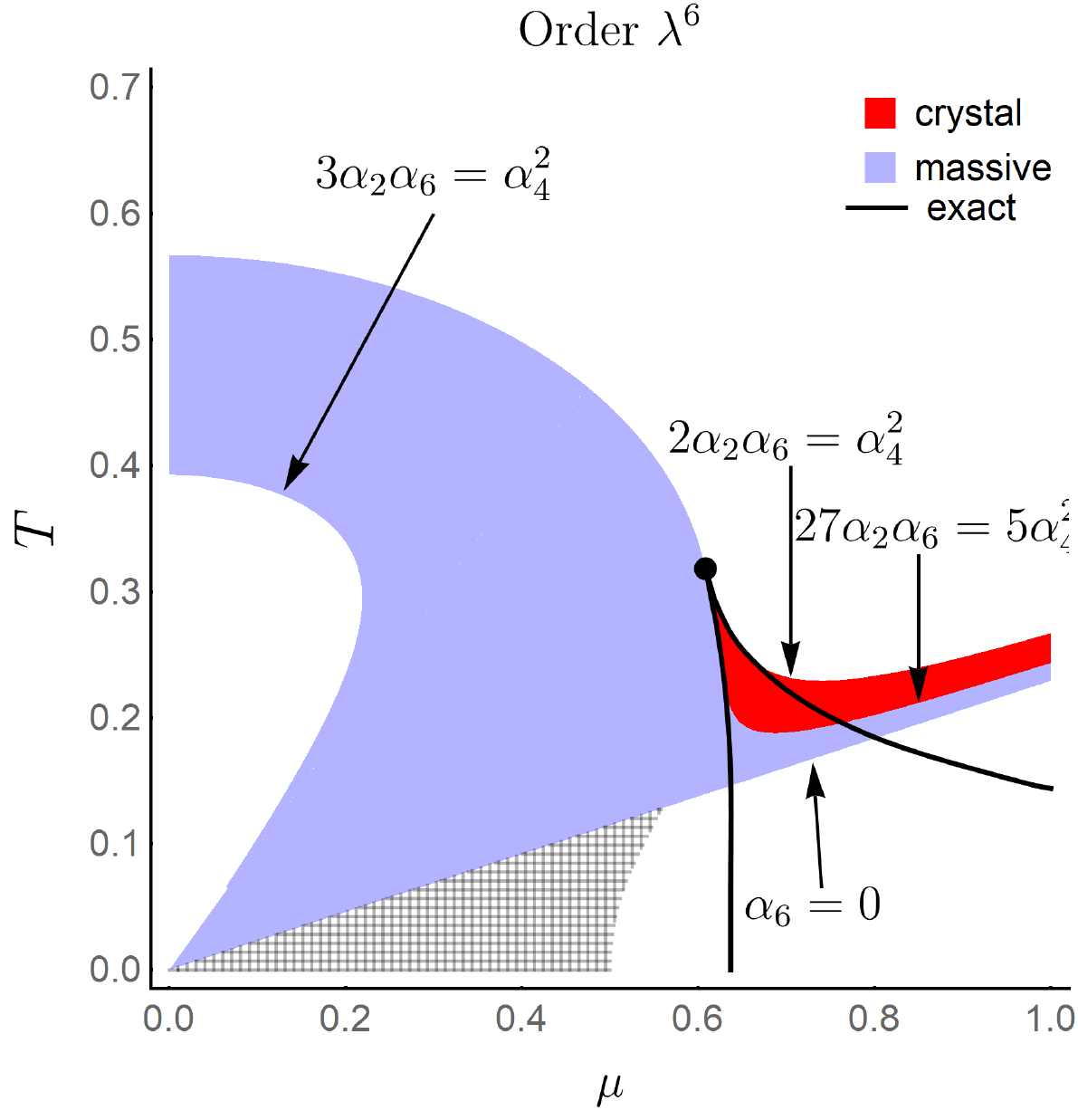}
		\caption{\textbf{Order $\lambda^6$ phase diagram}: Dot - tricritical point, white - massless phase, mesh -  analysis is invalid because $\Psi^{(6)}$ is unbounded from below. Solid line shows the boundaries of the exact numerically computed crystal phase. See Fig. \ref{fig:SixthOrderPhaseDiagramZoomed} for a close up view of the crystal phase.}
		\label{fig:SixthOrderPhaseDiagram}	
\end{figure}
Applying the minimization procedure to Ginzburg-Landau expansion \eqref{eq:GLExpAlln} truncated at order $\lambda^6$ ($N=3$ in \eqref{eq:truncatedGCP}) produces the coupled equations
\begin{equation} 
	\begin{aligned} \label{eq:alpha6min}
		\partial_\lambda \Psi^{(6)}&= 0 \implies 2 \lambda\left( \alpha_2 f_2 + 2 \lambda^2 \alpha_4 f_4 + 3 \lambda^4 \alpha_6 f_6 \right) = 0, \\
		\partial_\nu \Psi^{(6)} &= 0 \implies \lambda^2 \left(\alpha_2 f_2' + \lambda^2 \alpha_4 f_4' + \lambda^4 \alpha_6 f'_6 \right)= 0.
	\end{aligned}
	\end{equation}
The trivial solution ($\lambda = 0$) is the massless homogeneous phase, which can be a minimum only in the region $\alpha_2(T,\mu) > 0$. For $\nu = 0$, the first equation is satisfied trivially, while the second gives two solutions for $\lambda^2$. Both of these solutions are saddle solutions and not minima. For $\nu \neq 0$, these equations can be combined into a single equation which is linear in $\lambda^2$, the solution to which does minimize $\Psi^{(6)}$. This solution is the crystal phase, which is absent in lower order phase diagrams. The crystal phase breaks translational invariance and is characterized by a non zero scale parameter $\lambda$ and  elliptic parameter $\nu$ between 0 and 1. It is described by the equations
\begin{equation} \label{eq:O6MinValLambda}
\lambda  = \sqrt{- \left( \frac{3f_6f_2' - f_2f_6'}{3f_6f_4' - 2f_4f_6'} \right) \frac{\alpha_2}{\alpha_4}}, 
		\end{equation}
\begin{equation} \label{eq:O6MinValNu}
\alpha_2 \alpha_6 = \frac{\left(2 f_4 f_2' - f_2 f_4' \right)\left(3 f_6 f_4' - 2 f_4 f_6'\right) }{\left( f_2 f_6' - 3 f_6 f_2'\right)^2} \alpha_4^2,
\end{equation}
and is constrained to the region $\alpha_2(T,\mu) > 0$, $\alpha_4(T,\mu) < 0$. This constraint follows from the requirements that the Hessian matrix (\ref{eq:Hessian}) evaluated on the solution \eqref{eq:O6MinValLambda},\eqref{eq:O6MinValNu} is positive definite and that the minimizing $\lambda^2$ is non-negative. See Fig. \ref{fig:SixthOrderPhaseDiagram} for the phase diagram. 
\begin{figure} [!t]
		\raggedright \includegraphics[scale=0.7]{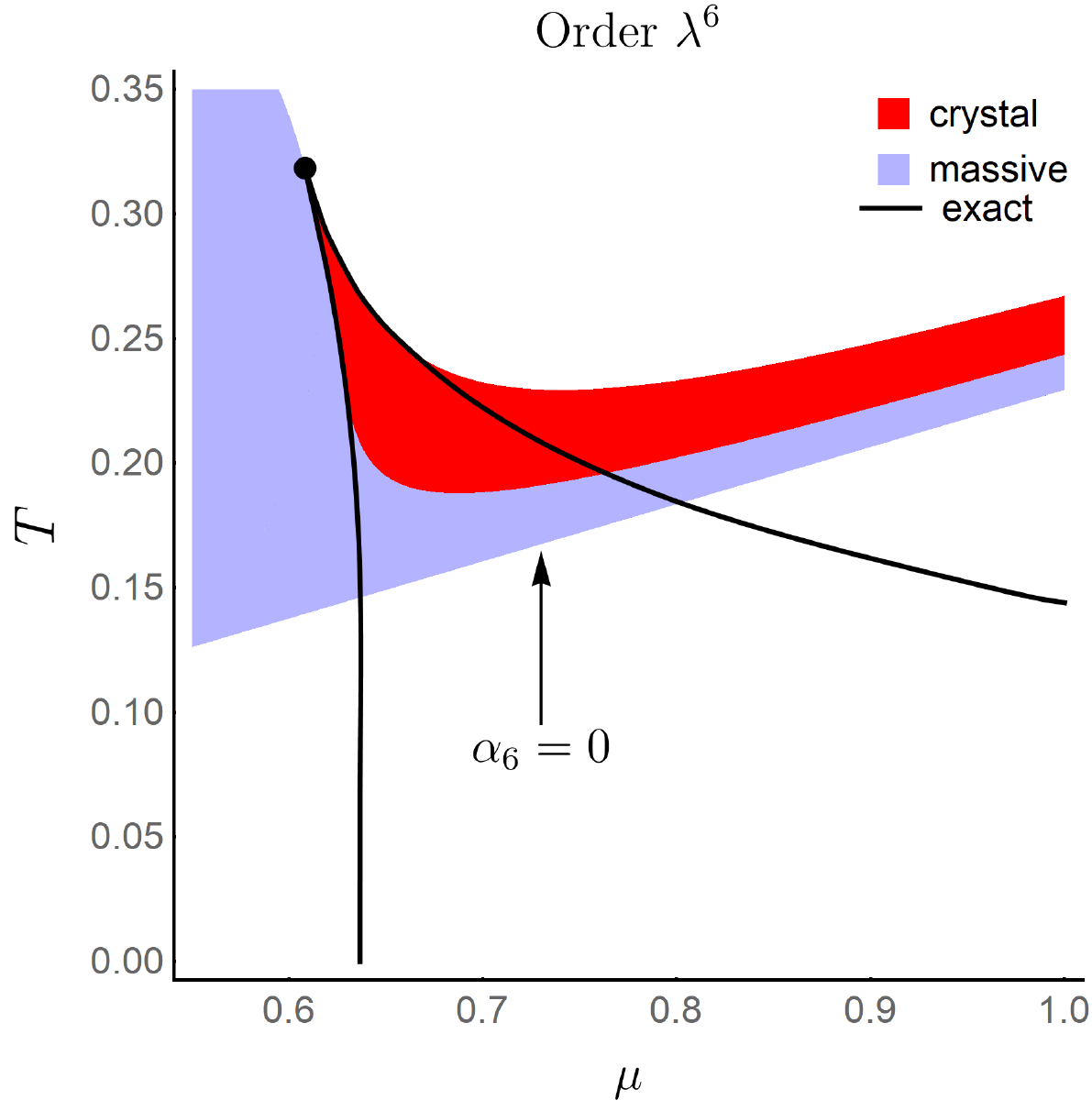}
		\caption{\textbf{Order  $\lambda^6$ crystal phase}:  Dot - tricritical point, white - massless phase. Solid line - exact numerically calculated boundary of the crystal phase. Compare with higher order phase diagrams in Figure \ref{fig:EighthOrderPhaseDiagram}.}
		\label{fig:SixthOrderPhaseDiagramZoomed}	
\end{figure}
Equation \eqref{eq:O6MinValNu} is an implicit equation for $\nu = \nu(T,\mu)$, which can be thought of as describing an infinite family of curves inside the crystal phase, each curve associated with a particular value of $\nu$. These curves span the entire crystal phase. In the limit $\nu \rightarrow 0$ equation \eqref{eq:O6MinValNu} reduces to that of the phase boundary between massless and crystal phases
	\begin{equation}
		\alpha_2(T,\mu) \alpha_6(T,\mu) = \frac{1}{2} \left(\alpha_4(T,\mu)\right)^2,
	\end{equation}
while in the limit $\nu \rightarrow 1$ it reduces to that of the phase boundary between crystal and massive homogeneous phases
\begin{equation}
		\alpha_2(T,\mu) \alpha_6(T,\mu) = \frac{5}{27} \left(\alpha_4(T,\mu)\right)^2.
\end{equation}
 The phase boundary betweeen massless and massive homogeneous phases is same as before, $\alpha_2(T,\mu) = 0,\, \alpha_4(T,\mu) > 0$. All three phases meet at a tricritical point given by the intersection of the massive-massless phase boundary $\alpha_2(T,\mu) =0$ and the crystal phase \eqref{eq:O6MinValNu}:
\begin{equation} \label{eq:tricrit}
\alpha_2(T,\mu) = 0 = \alpha_4(T,\mu).
\end{equation}
This result agrees perfectly with that from the exact solution. Solving \eqref{eq:tricrit} numerically gives the tricritical temperature and chemical potential:
\begin{equation}
		T_\text{tri-crit}  \approx 0.318,  \qquad		\mu_\text{tri-crit}  \approx 0.608
\end{equation}

The sliver of crystal phase near the tricritical point agrees very well with the exact crystal phase, but then as $\mu$ (or $T$) increases it diverges away, running asymptotic to the lower of the two lines on which $\alpha_6(T,\mu)$ vanishes. See Fig. \ref{fig:SixthOrderPhaseDiagram}.  

A final feature to note here is the existence of a spurious massless phase inside the massive homogeneous phase, whose boundary is given by $3 \alpha_2(T,\mu) \alpha_6(T,\mu) = \left(\alpha_4(T,\mu) \right)^2$. This is in disagreement with the exact phase diagram. As mentioned before, the small $\lambda$ expansion \eqref{eq:GLExpAlln} is expected to return increasingly accurate results as the exact massless ($\lambda=0$) boundary is approached, and artifacts such as this spurious massless phase may show up away from the exact massless phase boundary. For this reason, analysis of higher order phase diagrams will be restricted to the crystal phase, as it is proximate to the massless phase.

\subsection{Phase diagrams to orders $\lambda^8$ and $\lambda^{10}$}
\begin{figure*} [!t]
	\raggedright	\includegraphics[scale=0.7]{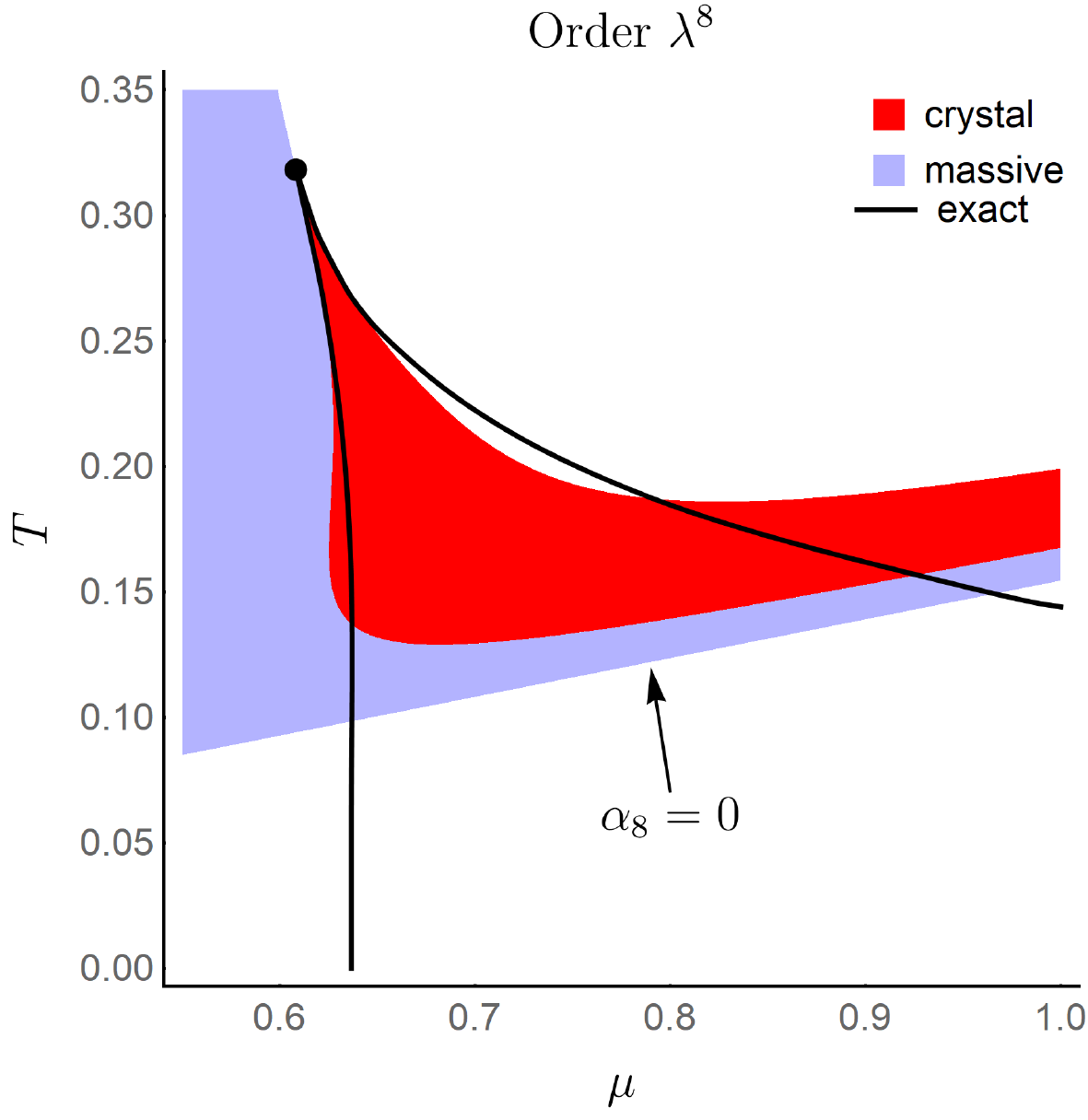} \includegraphics[scale=0.7]{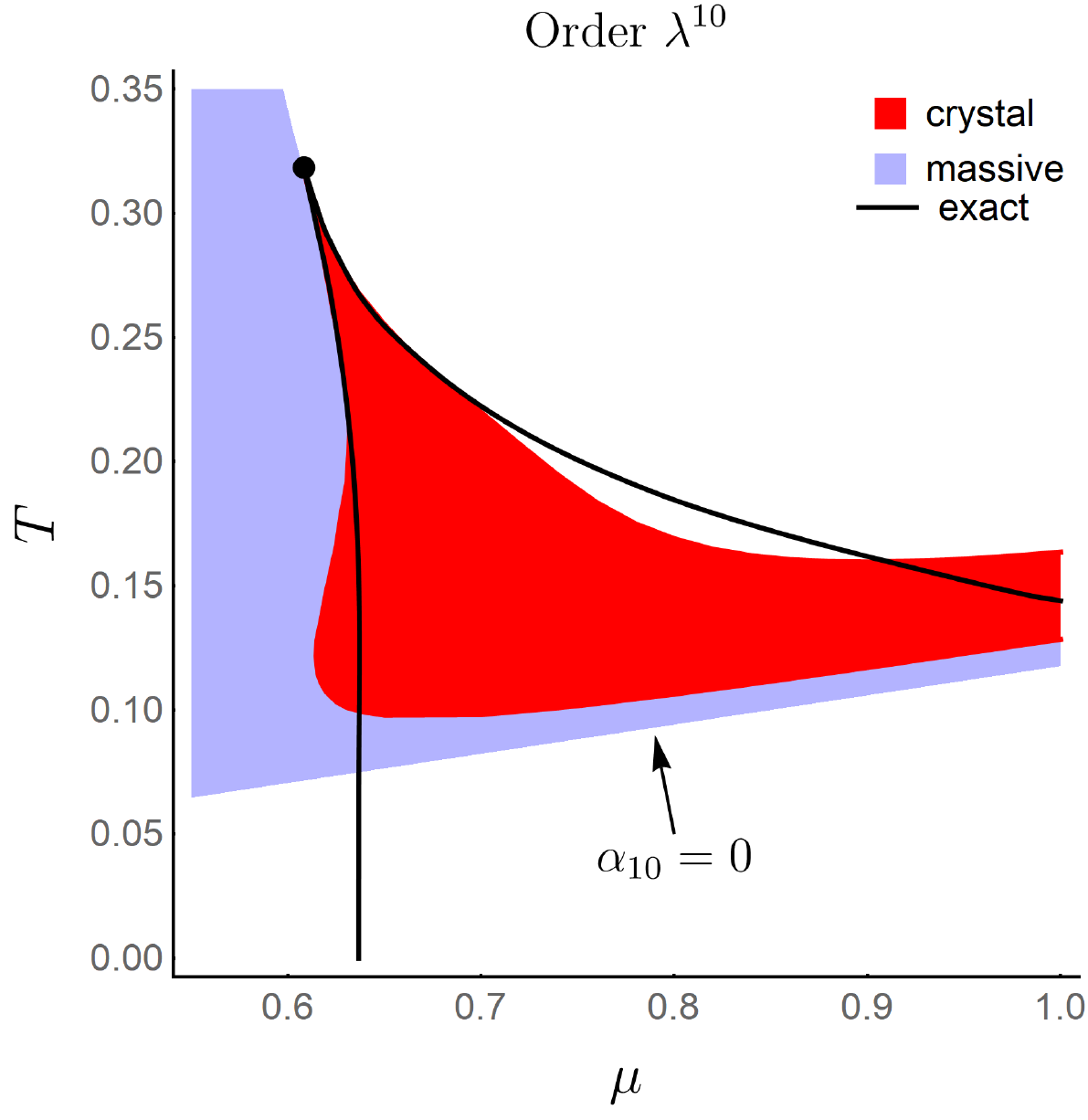}
		\caption{\textbf{Order $\lambda^8$ and $\lambda^{10}$ crystal phases}:  Dot - tricritical point, white - massless phase. Solid line - exact numerically calculated boundary of the crystal phase.}
		\label{fig:EighthOrderPhaseDiagram}	
\end{figure*}
Increasing the order of expansion to $\lambda^8$ improves the resemblance of the crystal phase to the exact crystal phase. The phase becomes broader, dips closer to the $\mu$-axis and agrees with the exact crystal phase further away from the tricritical point than the order $\lambda^6$ crystal phase does. See Fig. \ref{fig:EighthOrderPhaseDiagram} for the phase diagram.

At this order the minimization conditions (\ref{eq:minCond}) reduce to a quadratic equation in $\lambda^2$. One of the two solutions is actually a maximum, and must be rejected. The other solution is a minimum and gives the crystal phase. The resulting equation of the crystal phase is too cumbersome to be given here. The equations describing the phase boundaries between massless and crystal phases, and crystal and massive phases boundaries are simpler, and can be obtained by taking the limits $\nu \rightarrow 0$ and $\nu \rightarrow 1$ respectively.  
The massless-crystal phase boundary is given by 
\begin{equation}
	\alpha_8  = \frac{5}{54 \alpha_2^2}  \bigg(9 \alpha_2 \alpha_4 \alpha_6 - 4 \alpha_4^3  -\sqrt{2} \left( 2 \alpha_4^2 - 3 \alpha_2 \alpha_6 \right)^\frac{3}{2}\bigg),
\end{equation}
and the crystal-massive phase boundary is given by
\begin{equation}
\begin{aligned}
\alpha_8 = \frac{189 \alpha_6^2 \left(10 \alpha_4^2 - 27 \alpha_2 \alpha_6+ \alpha_4 \sqrt{100 \alpha_4^2 - 405 \alpha_2 \alpha_6}\,\right)}{2 \left(10 \alpha_4 + \sqrt{100 \alpha_4^2 - 405 \alpha_2 \alpha_6}\, \right)^3},
\end{aligned}
\end{equation}
both constrained to the region $\alpha_2(T,\mu) > 0$, $\alpha_4(T,\mu) < 0$. 
\begin{figure} [!t]
	\raggedright				
		\includegraphics[scale=0.72]{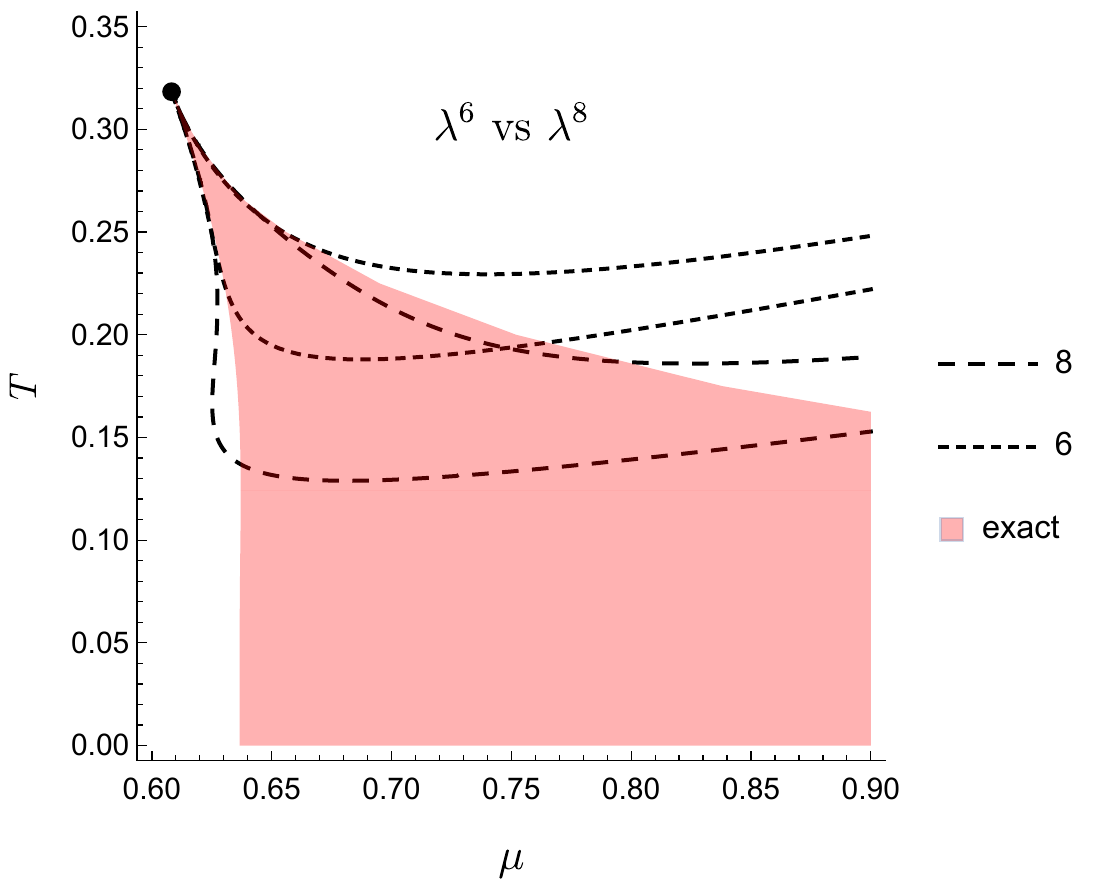}
		\includegraphics[scale=0.72]{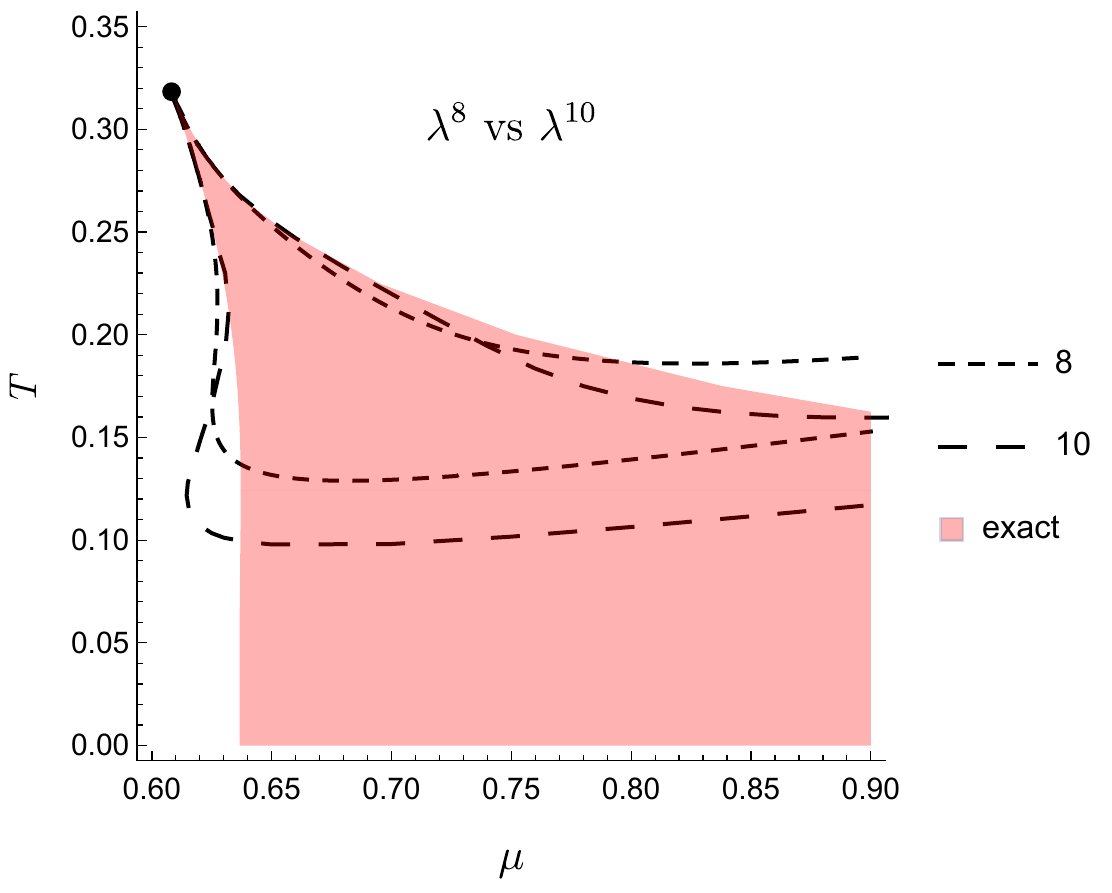}
		\caption{Comparison of crystal phases from third order expansion (outlined by small dashes) and fourth order expansion (outlined by longer dashes). Shaded region is the exact crystal phase.}
		\label{fig:Comparison3vs4}	
\end{figure}
The phase diagram from order $\lambda^{10}$  expansion has an even larger crystal phase, and is accurate to the true crystal phase even further from the tri-critical point. See Fig. (\ref{fig:EighthOrderPhaseDiagram}) for the phase diagram. The minimization equations (\ref{eq:minCond}) at this order reduce to a cubic equation in $\lambda^2$. Only one of the three solutions is a minimum. The equation of the crystal region is quite cumbersome at this order, and is left out.

For a quick comparison of the crystal phases at different orders and the exact crystal phase, see Fig. \ref{fig:Comparison3vs4}. It is evident that increasing the order of the Ginzburg-Landau expansion significantly improves the accuracy of the crystal phase.

A final point to note is that the crystal region in each case is asymptotic to a straight line in the $\mu-T$ plane. For order $\lambda^8$ crystal phase, this line is the lowest of the lines given by $\alpha_8(T,\mu) = 0 $, while for order $\lambda^{10}$ crystal phase, it is given by $\alpha_{10}(T,\mu) = 0$. This asymptotic behavior in fact exists at all orders, and is discussed next along with some other features of phase diagrams at arbitrary orders.

\subsection{Large order phase diagrams}
This section explores certain properties of phase diagrams obtained from large order Ginzburg-Landau expansions. The first of these properties holds for all $N \geq 3$ and large $\mu$ (or equivalently, large $T$): If $ T = a_{2N,j}\, \mu$ are the $N-1$ lines satisfying $\alpha_{2N}(T,\mu) = 0$, labelled by $j=1,2,..,N-1$, then the crystal region in the phase diagram of a $\lambda^{2N}$ Ginzburg-Landau expansion will asymptote to that particular line for which the slope $a_{2N,j}$ is smallest.

Before proving this statement, it is instructive to do a short exercise to understand the asymptotic behavior of the order $\lambda^6$ crystal phase. The equation of the crystal phase at order $\lambda^6$ is given by \eqref{eq:O6MinValNu}: $\alpha_2 \alpha_6 = I(\nu) \alpha_4^2$, or

\begin{equation} \label{eq:O6explicit}
 \left(\psi_{r,0} \left(\frac{\mu}{T}\right) + \ln (4\pi T) \right)\psi_{r,4}  \left(\frac{\mu}{ T}\right) = 3 I(\nu) \psi_{r,2}^2 \left(\frac{\mu}{T}\right)
\end{equation}
where 
\begin{equation}
 \begin{aligned}
 \psi_{r,n}(x) \equiv \text{Re } \psi^{(n)}\left(\frac{1}{2} + i\frac{x}{2 \pi} \right)\\
 \psi_{i,n}(x) \equiv \text{Im } \psi^{(n)}\left(\frac{1}{2} + i\frac{x}{2 \pi} \right)
 \end{aligned}
\end{equation}
and the function $I(\nu)$ is defined to capture the $\nu$ dependence in \eqref{eq:O6MinValNu} 
\begin{equation}
	I(\nu) =  \frac{\left(2 f_4 f_2' - f_2 f_4' \right)\left(3 f_6 f_4' - 2 f_4 f_6'\right) }{\left( f_2 f_6' - 3 f_6 f_2'\right)^2}.
\end{equation}
 Suppose $\mu = a \, T$ satisfies $\alpha_6(T,\mu) = 0 \iff \psi_{r,4}(\frac{\mu}{T}) = 0$. Then, in order to explore the crystal phase in the vicinity of the line $\alpha_6(T,\mu) = 0$, write
\begin{equation} \label{eq:ansatz}
 a T = \mu ( 1+ \epsilon(\mu))
\end{equation}
and expand \eqref{eq:O6explicit} about $\epsilon = 0$, keeping only the leading terms:
\begin{equation}
\begin{aligned}
	\left( \psi_0 \left(a\right) + \ln \frac{4\pi\mu}{a} \right) \frac{a \epsilon(\mu)}{2\pi} \psi_{i,5}\left( a \right) = 3 I(\nu) \psi_2 \left(a\right) \\
	\times \left( \psi_2 \left(a\right) +  \frac{a \epsilon(\mu)}{\pi}   \psi_{i,3}\left( a \right) \right).
\end{aligned}
\end{equation}
This equation can be solved for $\epsilon(\mu)$ to give
\begin{equation}
	\epsilon(\mu) = \frac{2\pi}{a} \frac{3 I(\nu) \left(\psi_2 \left(a\right) \right)^2  }{\left( \psi_0 \left(a\right) + \ln \frac{4\pi\mu}{a} \right) \psi_{i,5}\left( a \right)  - 6 I(\nu) \psi_2 \left(a\right) \psi_{i,3}\left( a \right)  } 
\end{equation}
which, for large $\mu$, implies that $\epsilon(\mu)$ is a small quantity and scales as
\begin{equation} \label{eq:epsilonscaling}
 	\epsilon(\mu) \sim \frac{1}{\ln \mu}.
 \end{equation}
This is consistent with the initial assumption that, for large $\mu$, equation \eqref{eq:O6explicit} can be expanded about $\epsilon = 0$. Further, the scaling \eqref{eq:epsilonscaling} implies that the crystal phase is asymptotic to the line $\alpha_6(T,\mu) = 0$. 

 The equation for the minimizing $\lambda$ is given by \eqref{eq:O6MinValLambda}:
\begin{equation}
	\lambda = \sqrt{-H(\nu) \frac{\alpha_2(T,\mu)}{\alpha_4(T,\mu)}} = \sqrt{-H(\nu)I(\nu) \frac{\alpha_4(T,\mu)}{\alpha_6(T,\mu)}}
\end{equation}
where 
\begin{equation}
	H(\nu) = \frac{3f_6f_2' - f_2f_6'}{3f_6f_4' - 2f_4f_6'}.
\end{equation}
This expression for the minimizing $\lambda$ in terms of $\alpha_4$ and $\alpha_6$  implies that for large $\mu$ and in the vicinity of the line $\alpha_6(T,\mu) = 0$ (i.e. $\mu \rightarrow \infty$ and $a T = \mu ( 1+ \epsilon(\mu))$) $\lambda$ scales as
\begin{equation} \label{eq:O6lambdascaling}
	\frac{\lambda}{T} \sim \frac{1}{\sqrt{\epsilon(\mu)}} \implies \frac{\lambda}{T} \sim \sqrt{\ln \mu}
\end{equation}
\quad\\
It turns out that, for $N \geq 3$, $\dfrac{\lambda}{T}$ always scales as $\dfrac{1}{\sqrt{\epsilon(\mu)}}$, as will be seen later.

The asymptotic behavior of order $\lambda^{2N}, N \geq 3,$ crystal phase can be analysed in a manner similar to that of order $\lambda^6$. The only difference is that unlike order $\lambda^6$ crystal phase, an explicit equation is unknown for order $\lambda^{2N}$ crystal phase. Thus the only option is to work with the equations
\begin{equation} \label{eq:MinCoupled}
	\begin{aligned}
		&\partial_\lambda \Psi^{2N} = 0 \implies \sum_{n=1}^N n \alpha_{2n} f_{2n} \lambda^{2n-2} = 0, \\
		&\partial_\nu \Psi^{2N} = 0 \implies \sum_{n=1}^N \alpha_{2n} f'_{2n} \lambda^{2n-2} = 0.
	\end{aligned}
\end{equation}
Suppose $1/a_N$ is the slope of the lowest line satisfying $\alpha_{2N}(T,\mu) = 0$, i.e. $a_N$ is the smallest number which satisfies $\psi_{r,2N-2}(a_N) = 0$. The crystal phase in the vicinity of this line is explored by writing $a_N T = \mu (1 + \epsilon_N(\mu,\nu))$ and expanding the equations \eqref{eq:MinCoupled} about $\epsilon_N = 0$. 
\begin{figure*} [!t]			
		\raggedright \includegraphics[scale=0.37]{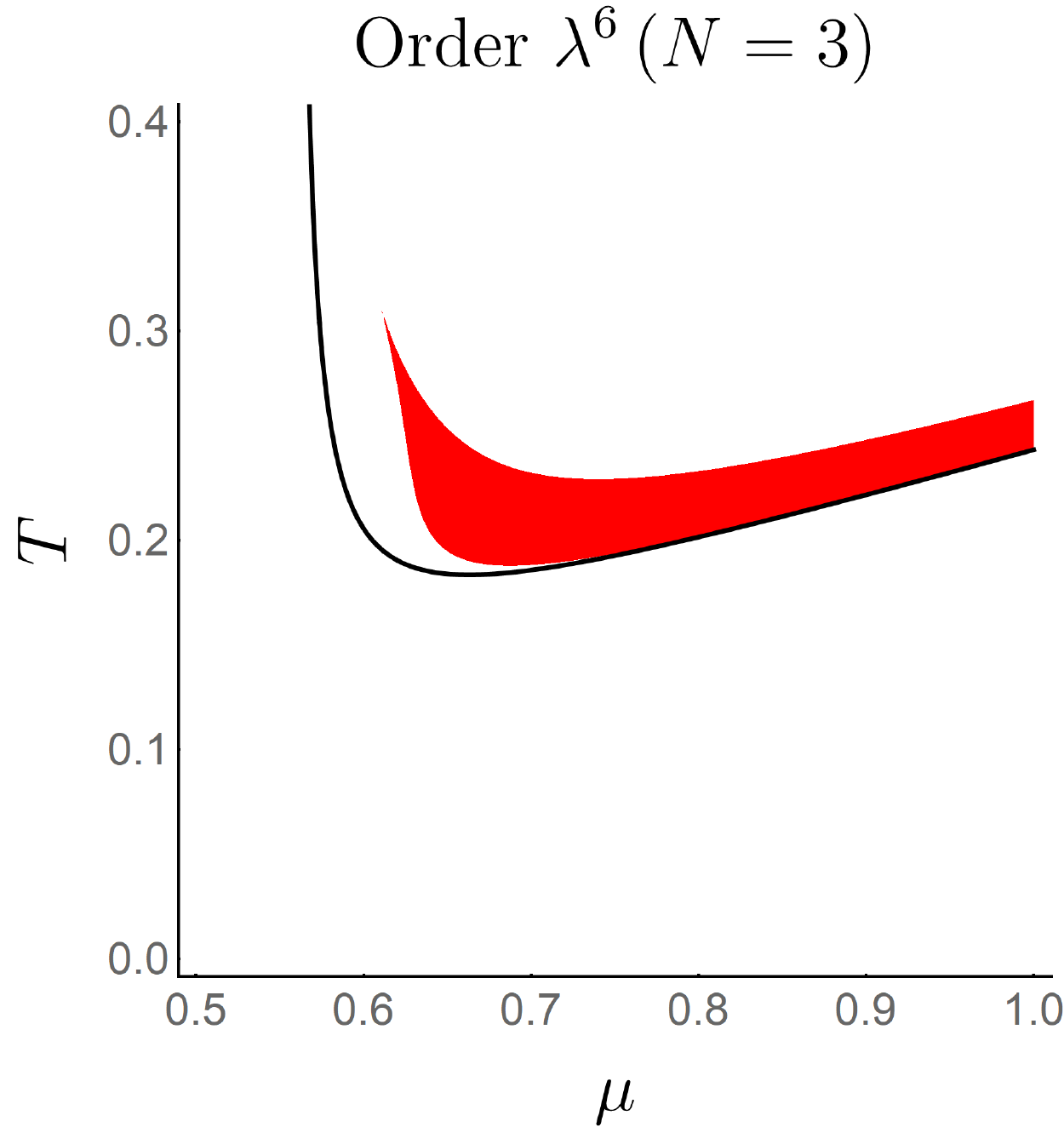} \includegraphics[scale=0.37]{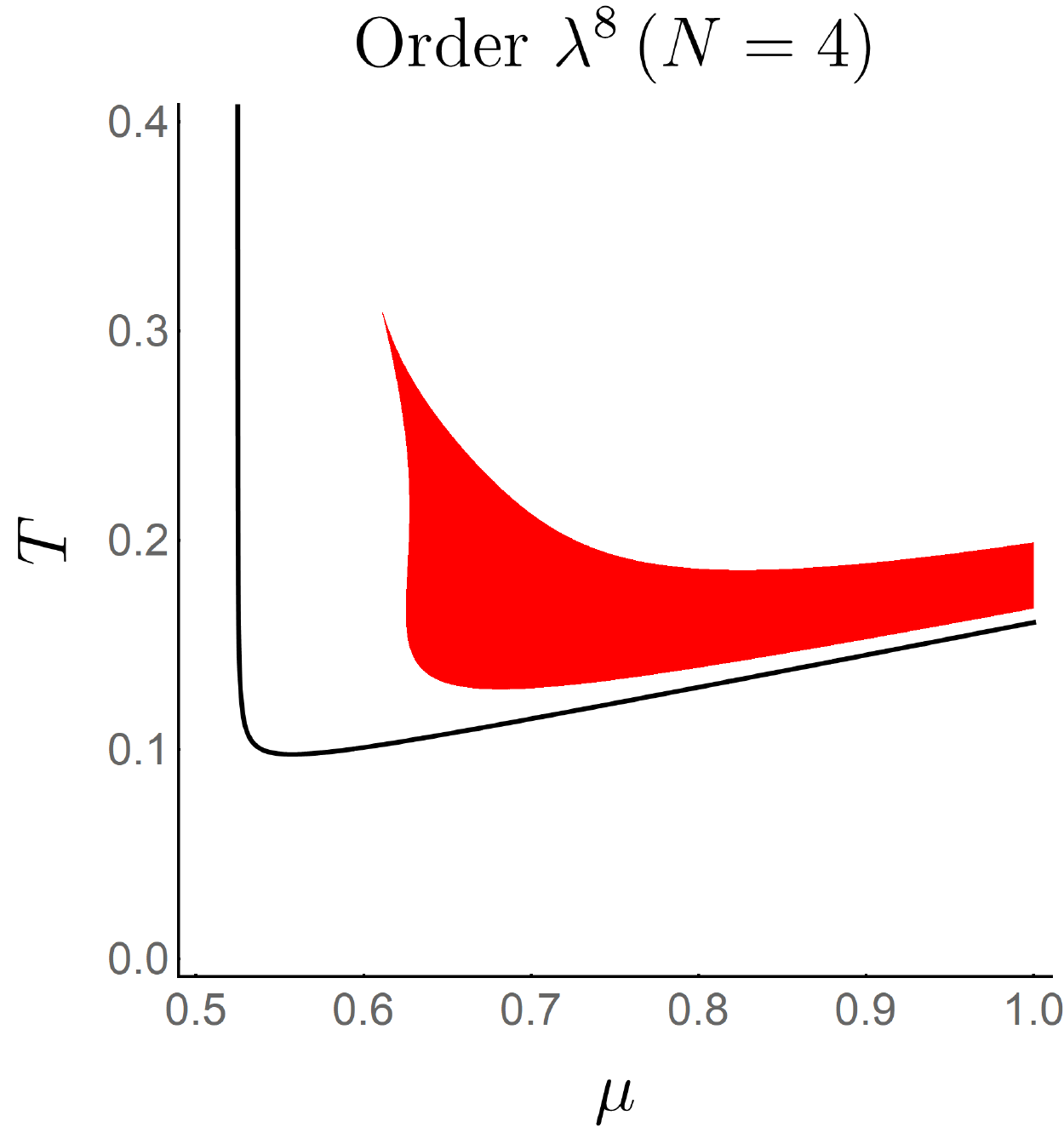} 	\includegraphics[scale=0.37]{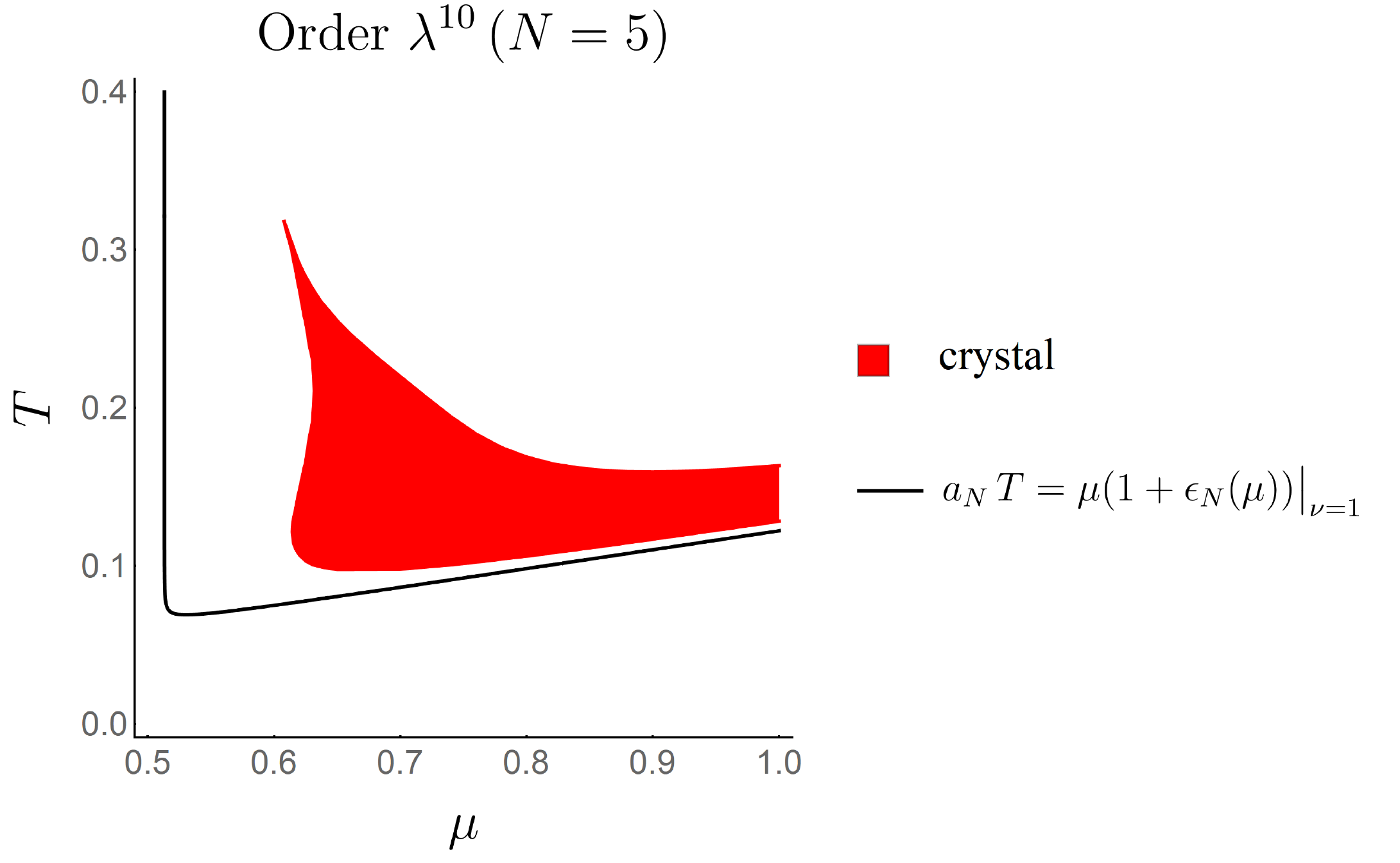}
		\caption{Comparison of the crystal phase from Ginzburg-Landau expansions at three different orders ($N=3,4,5$) to the leading asymptotic approximation to the equation of lower edge of the crystal phase $a_N T = \mu \left( 1 + \epsilon_N(\mu) \right)|_{\nu = 1}$.}
		\label{fig:asympCompare}
\end{figure*}
The $T$-dependent functions in the expansion expand as:
\begin{align}
	& \ln (4 \pi T) \approx \ln \frac{4 \pi \mu}{a} + \epsilon \quad  \implies \alpha_2 \sim \ln \mu, \\
	& \psi_{r,n}\left(\frac{\mu}{T}\right) \approx  \psi_{r,n}(a) + \frac{a \epsilon}{2\pi}  \psi_{i,n+1}(a), \quad n \neq 2N-2 \\
	& \psi_{r,2N-2}\left(\frac{\mu}{T}\right) \approx  \frac{a \epsilon}{2\pi}  \psi_{i,2N-1}(a).
\end{align}
The $N$ and $\nu$ dependencies in $a_N$ and $\epsilon_N(\mu,\nu)$ have been supressed in the above equations. Eliminate $\alpha_2$ to obtain
\begin{equation}	\label{eq:MinCondAnyN}
		\sum_{n=2}^N (n f'_2 f_{2n} - f_2  f'_{2n} ) \alpha_{2n} \lambda^{2n-2} = 0
\end{equation}
It is useful to rewrite this equation in terms of the dimensionless quantities
\begin{equation} \label{eq:dimenQuant}
 \hat{\alpha}_{2n}\left(\frac{\mu}{T}\right) = T^{2n-2} \alpha_{2n}(T,\mu)\, , \qquad m = \frac{\lambda^2}{T^2}.
\end{equation}
Also, define the functions $g_n(\nu)$ to collect all $\nu$ dependence:
\begin{equation}
	g_n(\nu) = n f'_2 f_{2n} - f_2  f'_{2n}.
\end{equation}
The functions $g_n(\nu)$ are monotonic increasing non-negative for all $n \geq 2$. After this notational change \eqref{eq:MinCondAnyN} looks like
\begin{equation}	\label{eq:MinCondAnyNDimLess}
		\sum_{n=2}^N g_n \hat{\alpha}_{2n}\left(\frac{\mu}{T}\right) m^{n-1} = 0.
\end{equation}
This equation implies that $m$ is a function of the ratio $\mu/T$, and there is no other $\mu$ or $T$ dependence in $m$. In the vicinity of the line $\alpha_{2N} (T,\mu) = 0$, i.e. for $\hat{\alpha}_{2N} \sim \epsilon(\mu)$, one of the solutions is given by
\begin{equation} \label{eq:largeMuM}
	m = - \dfrac{g_{N-1} \, \hat{\alpha}_{2N-2} \left(\frac{\mu}{T}\right)}{g_N \, \hat{\alpha}_{2N} \left(\frac{\mu}{T}\right)} \sim \frac{1}{\epsilon(\mu)}
\end{equation}
This solution is guaranteed to be positive (required for a real $\lambda$) on one side of line $\hat{\alpha}_{2N}\left(\mu/T\right) = 0$, the side on which $\hat{\alpha}_{2N}\left(\mu/T\right)$ is positive. This is true because $\hat{\alpha}_{2N-2}(\mu/T)$ is always negative on the line $\hat{\alpha}_{2N}\left(\mu/T\right) = 0$ and in the immediate vicinity of it. In fact a stronger statement is true: If $a$ is the largest solution of $\hat{\alpha}_{2N}(x) = 0$, then
\begin{equation} \label{eq:magic}
	\hat{\alpha}_{2N-2 k} (a \pm \delta) < 0, \text{ for } 0 \leq \delta \ll 1, \text{ and } k = 1,2,..,N-2
\end{equation}
The other $N-3$ solutions of \eqref{eq:MinCondAnyNDimLess} are, to leading order in $\epsilon(\mu)$, the roots of the polynomial $\sum_{n=2}^{N-1} g_n \hat{\alpha}_{2n} m^{n-1}$. These are either non-positive or complex, because of \eqref{eq:magic}, and must be rejected. Thus, there is only one physical solution of equation \eqref{eq:MinCondAnyN} and it scales as
\begin{equation}
\frac{\lambda}{T} = \sqrt{m}  \sim \frac{1}{\sqrt{\epsilon(\mu)}}.
\end{equation}
Next, eliminate $\alpha_{2N}$ from the coupled equations \eqref{eq:MinCoupled}
\begin{equation}	\label{eq:MinCondAnyNWithoutAlpha2N}
		\sum_{n=1}^{N-1} \left(\frac{n f_{2n}}{N f_{2N}} - \frac{f_{2n}'}{f_{2N}'} \right) \hat{\alpha}_{2n}\left(\frac{\mu}{T}\right) m^{n-1} = 0.
\end{equation}
Plugging the solution \eqref{eq:largeMuM} into this equation and keeping only the leading power of $\epsilon(\mu)$ gives 
\begin{equation} \label{eq:epsilonScaling}
\begin{aligned}
		\bigg(\frac{f_{2}}{N f_{2N}} & - \frac{f_{2}'}{f_{2N}'} \big) \hat{\alpha}_{2} + \bigg(\frac{(N-1) f_{2N-2}}{N f_{2N}} \\
		& - \frac{f_{2N-2}'}{f_{2N}'} \bigg) \hat{\alpha}_{2N-2} \left(- \dfrac{g_{N-1} \hat{\alpha}_{2N-2}}{g_N\hat{\alpha}_{2N}}\right)^{N-2} \approx 0\\
		\implies \quad & \epsilon_N(\mu) \sim \left(\frac{1}{\ln \mu}\right)^{\frac{1}{N-2}}
\end{aligned}
\end{equation}
This proves the asymptotic nature of the crystal phase at order $\lambda^{2N}$ to the line $\alpha_{2N}(T,\mu) = 0$. This result agrees with the explicit calculation of $\epsilon(\mu)$ from equations of crystal phases at orders $\lambda^6$ ($N=3$), equation \eqref{eq:O6lambdascaling},  and $\lambda^8$ ($N=4$, not included here).

The above result leads to the second important property of phase diagrams for $N \geq 3$: As $N$ increases, $a_{2N}$ decreases (see Fig. \ref{fig:SignsAlpha}), which implies that increasing the order of the Ginzburg-Landau expansion will cause the crystal phase to dip lower and closer to the $\mu$ axis, making it increasingly accurate. As a side effect, the crystal phase can never touch the $\mu$-axis as $T=0$ is not a solution of $\alpha_{2N}(T,\mu) = 0$ for any finite $\mu$ and $N \geq 2$. This, in hindsight, is to be expected as the expansion \eqref{eq:GLExpAlln} can be thought of as an expansion in inverse powers of $T$. As one probes closer and closer to $T=0$, more and more terms are needed in the Ginzburg-Landau expansion to maintain accuracy. At exactly $T=0$, one would need all terms in the expansion.

Equation \ref{eq:epsilonScaling} allows one to write an approximate expression for the crystal phase for any given value of $\nu$, in the form $a_N T = \mu ( 1 + \epsilon_N(\mu,\nu))$. For example, for the lower edge of the crystal region, $\nu$ is set to 1, and $\epsilon_N(\mu)$ takes the form
\begin{equation}
\begin{aligned}
& \epsilon_N(\mu)\big|_{\nu=1} = \frac{2 \pi}{a_N} \left|\frac{\psi_{r,2N-4}(a_N)}{\psi_{i,2N-1}(a_N)} \right|  \frac{(N-1)(N-2)(2N-1)}{2N-3} \\
& \times \left(\frac{|\psi_{r,2N-4}(a_N)|}{(N-1)! (N-2)! (2N-3)}   \frac{1}{\ln \frac{4\pi \mu}{a_N} + \psi_{r,0}(a_N)}\right)^\frac{1}{N-2} 
\end{aligned}
\end{equation}
Of course, this is just the leading (asymptotic) behavior of the crystal phase and its accuracy increases with $\mu$. See Fig. \ref{fig:asympCompare}. The $\mu$-dependence in sub-leading corrections is generically
\begin{equation}
	 \left( \frac{1}{\ln (4\pi \mu) - \ln a_N + \psi_{r,0}(a_N)}\right)^k, \quad k > \frac{1}{N-2} ,
\end{equation}
but these are beyond the scope of this paper.

Finally, the large $N$ limit of $\epsilon_N(\mu)|_{\nu=1}$ is
\begin{equation} \label{eq:epsilonN}
\begin{aligned}
\epsilon_N(\mu)|_{\nu=1} \xrightarrow{N \rightarrow \infty} & \frac{2 \pi e^2}{a_N} \left|\frac{\psi_{r,2N-4}(a_N)}{\psi_{i,2N-1}(a_N)} \right| \left( \frac{|\psi_{r,2N-4}(a_N)|}{\ln 2\mu}\right)^\frac{1}{N} \\ 
& \sim  \frac{1}{N}  \left( \frac{1}{\ln 2\mu}\right)^\frac{1}{N}
\end{aligned}
\end{equation}
For large $N$, $\epsilon_N(\mu)|_{\nu=1}$ is real and goes as $1/N$, as long as $\mu > 1/2$. Otherwise, it is either undefined ($\mu=1/2$) or complex ($\mu < 1/2$), suggesting that the lower edge of the crystal phase can never come close to $\mu$-axis for values of $\mu \leq 1/2$. This indicates that the phase transition from massive to crystal phase can occur only for $\mu > 1/2$. Since \eqref{eq:epsilonN} is just the leading term in an asymptotic approximation to the equation of the lower edge of the crystal phase, one would presumably need all the terms to exactly locate the $T=0$ critical chemical potential. However, it does agree with the fact that the known value of critical chemical potential ($\mu_c = 2/\pi$)  \cite{thiesOriginal,bdt} is greater than $1/2$.

\section*{Conclusion}
A Ginzburg-Landau type approach to thermodynamics of the Gross-Neveu model predicts several features of the exact phase diagram. Using certain asymptotical properties of the crystal phase, it can be shown that the Ginzburg-Landau crystal phase becomes more and more accurate as the order of the Ginzburg-Landau expansion increases.

\begin{acknowledgments}
I thank Gerald Dunne for the many helpful discussions. This work was supported by U.S. Department of Energy, Office of Science, Office of High Energy Physics under Award Number DE-SC0010339.
\end{acknowledgments}


\bibliography{gngl}
\bibliographystyle{unsrt}


\end{document}